\documentclass[amsmath,reprint,prb,longbibliography,superscriptaddress,bibnotes]{revtex4-2}
\usepackage{amsthm,amssymb,amsfonts,graphicx,verbatim, xcolor,bm} % The basics
\usepackage{hyperref} % URL links
\usepackage[utf8]{inputenc} % encoder
\usepackage{siunitx} % nice units
\usepackage{braket}
\usepackage{ulem}
\begin{document}

\title{Superlattice-Stabilized Fractional Chern Insulators in a ``Berry Basin" Model}
\author{Yongxin Zeng}
\email{yongxin.zeng@yale.edu}
\affiliation{Department of Physics, Columbia University, New York, NY 10027}
\affiliation{Department of Physics, Yale University, New Haven, CT 06520}
\author{Andrew J. Millis}
\affiliation{Department of Physics, Columbia University, New York, NY 10027}
\affiliation{Center for Computational Quantum Physics, Flatiron Institute, New York, NY 10010}

\begin{abstract}
The discovery of integer and fractional Chern insulators in rhombohedral graphene, following their discovery in the isolated topological flat band in twisted MoTe$_2$, motivates a search for the conditions under which an electronic band can host a topological correlated state. In this Letter we introduce a ``Berry basin" model which captures the essential features of rhombohedral graphene: a dispersive band with a flat bottom, momentum-space winding of a ``layer" degree of freedom, and Coulomb interactions, while also including periodic potentials, enabling controlled studies of the interplay of quantum geometry, superlattice modulation, and electron-electron interactions in producing correlated topological states beyond the isolated topological band paradigm. 
While integer Chern insulators emerge at strong coupling without a superlattice potential, stabilizing the fractional state requires nonzero potential for all interaction strengths we access. We provide insight into the greater robustness of integer and fractional Chern insulators in rhombohedral graphene when electrons are polarized away from the moir\'e interface. Our results establish a new class of systems that host robust fractional Chern insulators and provide a controlled numerical platform to test competing theories of interacting topological phases.
\end{abstract}

\maketitle

{\it Introduction.---}Fractional Chern insulators (FCIs) are topologically ordered electronic phases with fractionally quantized Hall response in the absence of external magnetic field. Theoretically proposed~\cite{sheng2011fractional, neupert2011fractional, tang2011high, sun2011nearly, regnault2011fractional} over a decade ago, FCIs have now been observed in twisted MoTe$_2$ bilayers~\cite{cai2023signatures, park2023observation, zeng2023thermodynamic, xu2023observation} and in hBN-encapsulated rhombohedral multilayer graphene~\cite{lu2024fractional, lu2025extended, xie2025tunable, waters2025chern, aronson2025displacement, choi2025superconductivity, butler2026fci}. These two platforms pose very different theoretical problems. In twisted MoTe$_2$, the moir\'e modulation acts as a periodic effective magnetic field~\cite{wu2019topological, morales2024magic, shi2024adiabatic} that produces a narrow, isolated Chern band~\cite{devakul2021magic, jia2024moire, zhang2024polarization}, and the FCI states follow from the conventional paradigm of partially filling the lowest Landau level. In rhombohedral graphene, this starting point fails since the moir\'e formed with the aligned hBN substrate does not produce an isolated topological flat band~\cite{herzog2024moire, park2023topological}, and the FCIs are observed to be more robust when the moir\'e acts on the layer {\it away} from the doped electrons and thus barely modifies the single-particle band structure.

Mean-field studies find robust interaction-induced Chern insulating states that survive the weak-moir\'e limit~\cite{dong2024anomalous, dong2024theory, zhou2024fractional, guo2024fractional, kwan2025moire} (anomalous Hall crystals~\cite{zeng2024sublattice, tan2024parent, soejima2024anomalous, dong2024stability, sheng2024quantum}), suggesting that the moir\'e serves mainly to pin an electron crystal and raising the possibility that FCI states may be considered as arising from hole-doping this parent state. This picture is supported by single-band-projected exact diagonalization (ED)~\cite{dong2024anomalous, dong2024theory, zhou2024fractional, guo2024fractional, huang2024self, huang2025fractional} and analytic constructions~\cite{tan2025variational, desrochers2026energetics, maymann2026skyrmion} but is inconsistent with the observed absence of integer and fractional Chern insulators in moir\'eless rhombohedral graphene~\cite{han2025signatures, seo2025family, han2026evidence} and is challenged by the collapse of FCIs in multiband ED~\cite{yu2025moire, li2025multi, regnault2026moire}. However, understanding the physics of moir\'e and band-mixing effects numerically has proven difficult because the absence of an isolated band makes ED calculations uncontrolled.

In this Letter we take a different route. We construct a minimal model---the ``Berry basin" model---that, while capturing the essential physics of rhombohedral graphene, can be generalized to explore the physics more broadly while remaining numerically tractable. The model features a basin-shaped single-particle band that is nearly dispersionless near its bottom and steeply dispersive outside the flat region and a two-component spinor Bloch wave function that exhibits nontrivial momentum-space winding. However, unlike the ``trashcan" model~\cite{bernevig2025berry} of rhombohedral graphene that lives in unbounded momentum space, our model also includes a strong periodic potential that defines a compact Brillouin zone (BZ) and isolates the basin-shaped band from remote bands, enabling controlled band-projected ED calculations. In addition, the model contains a commensurate ``moir\'e" superlattice potential with $\sqrt{3}\times\sqrt{3}$ periodicity (relative to the lattice of the strong periodic potential) that splits the band into three minibands that are generally not isolated from one another; nevertheless, we find it to be essential to the correlated physics at superlattice filling $\nu=1$ and $\nu=1/3$.

%A commensurate ``moir\'e" superlattice potential with $\sqrt{3}\times\sqrt{3}$ periodicity folds the band into a mini Brillouin zone (mBZ), splitting it into three minibands that are generally not isolated from one another. We perform ED at moir\'e fillings $\nu=1$ and $1/3$ and find that a honeycomb moir\'e potential $V_m$ stabilizes integer and fractional Chern insulator ground states over a broad range of interaction strength. While sufficiently strong interactions can stabilize integer Chern insulators at $V_m=0$---the commensurate counterpart of the anomalous Hall crystal~\cite{zeng2024sublattice, tan2024parent, soejima2024anomalous, dong2024stability, sheng2024quantum}, we do not find fractional anomalous Hall crystals at any interaction strength we access. Except in the weak-interaction regime, the critical moir\'e potential strength required to gap the interacting system is lower when it acts on the layer away from electrons at the band bottom, reproducing the counterintuitive phenomenology in rhombohedral graphene.

{\it Model.---}We consider spinless electrons with quartic dispersion
\begin{equation}
    \epsilon_k = \frac{k^4}{4} - \beta k^2,
\end{equation}
where the quadratic coefficient $\beta$ flattens the bottom of the band. We take the Bloch wave function to be a two-component spinor
\begin{equation}
    \ket{u_{\bm k}} = \begin{pmatrix}
        \cos\frac{\theta_k}{2} \\ e^{-iN\phi_{\bm k}} \sin\frac{\theta_k}{2}
    \end{pmatrix}
\end{equation}
whose azimuthal angle $\phi_{\bm k} = \arg(k_x+ik_y)$ winds $N$ times as the momentum is wound around the momentum space origin. This two-component spinor is the analog of the layer degree of freedom in spin- and valley-polarized rhombohedral graphene; we accordingly refer to the two components as the top and bottom ``layers". The polar angle $\theta_k$ parametrizes the layer content of the state at $\bm k$. We choose it to interpolate monotonically from $\theta_0=0$ to $\theta_{\infty}=\pi$ over a scale set by a parameter $\lambda$; explicitly, we take $\theta_k = \pi\tanh[(\lambda k/\pi)^2]$. Note that in the two-band model of rhombohedral graphene~\cite{min2008electronic} the winding number of Bloch spinors is related to the exponent characterizing the band dispersion, but here we define a more general model.

\begin{figure}
    \centering
    \includegraphics[width=\linewidth]{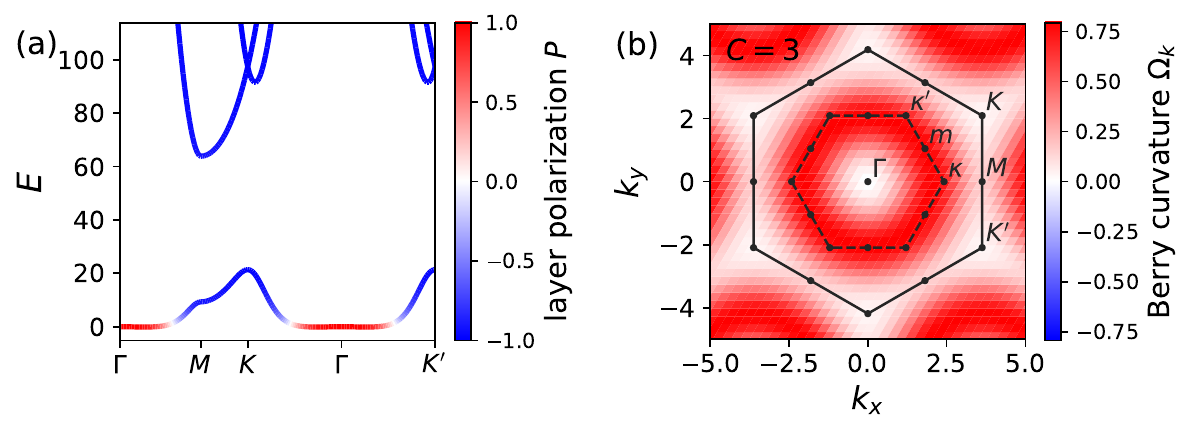}
    \includegraphics[width=\linewidth]{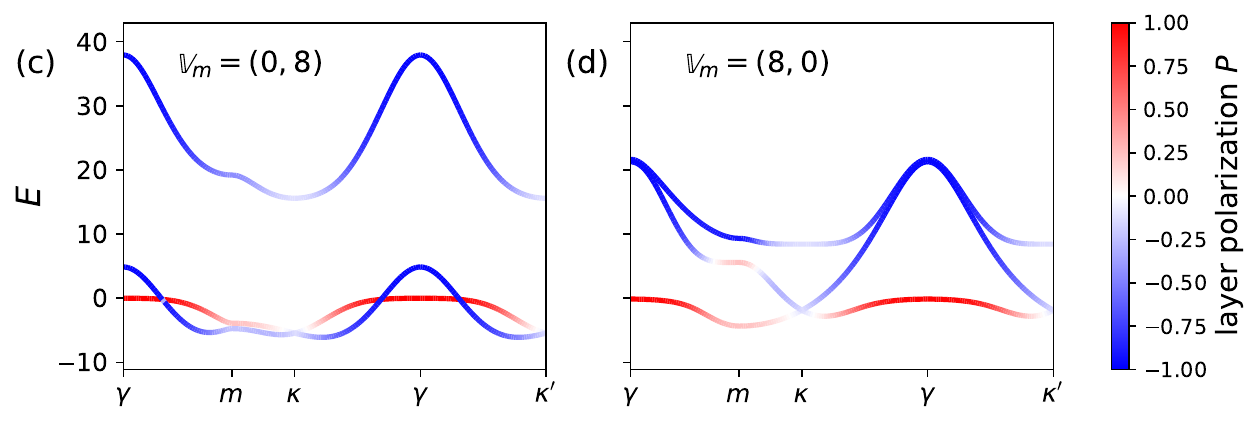}
    \caption{Single-particle bands of the Berry basin model with $N=3, \lambda=1, V_0=-25$, and $\beta=0.2$, the values used throughout this work. (a) Band structure in the absence of $V_m$. The color scale indicates the layer polarization $P$ defined such that $P=+1$ ($-1$) when the Bloch state is fully polarized to the top (bottom) layer. (b) Berry curvature $\Omega_{\bm k}$ of the lowest band in (a). (c) Minibands in the presence of a moir\'e superlattice potential $V_m=8$ acting on the bottom layer. (d) Same as (c) but with $V_m$ acting on the top layer.}
    \label{fig:bands}
\end{figure}

To compactify the band into a BZ, we add a periodic potential 
\begin{equation}
    V_0(\bm r) = \sum_{i=1}^3 V_0 e^{i\bm G_i \cdot \bm r} + {\rm c.c.}
\end{equation}
to the bottom layer. Here $\bm G_1 = (4\pi/\sqrt{3},0)$ and its $C_3$-partners $\bm G_{2,3}$ are the first-shell reciprocal lattice vectors of a triangular lattice with lattice constant $a=1$. Fig.~\ref{fig:bands}(a) shows the resulting band structure for $V_0 = -25, N=3, \lambda=1, \beta=0.2$, where a large gap separates the lowest basin-shaped band from the remote bands. These four parameters are held fixed throughout this work. The layer polarization indicated by the color scale shows that the Bloch states at the band bottom are strongly polarized to the top layer while those in the dispersive region and in the remote bands are polarized to the bottom layer. Fig.~\ref{fig:bands}(b) plots the Berry curvature of the lowest band whose Chern number is equal to the winding number $N=3$.

On top of the periodic potential $V_0(\bm r)$, we add a superlattice (``moir\'e") potential 
\begin{equation}
    V_m(\bm r) = \sum_{i=1}^3 V_m e^{i\bm g_i\cdot \bm r} + {\rm c.c.},
\end{equation}
where $\bm g_1 = (0,4\pi/3)$ and its $C_3$-partners $\bm g_{2,3}$ are the first-shell reciprocal lattice vectors of a commensurate $\sqrt{3}\times\sqrt{3}$ superlattice imposed on the parent triangular lattice. We write $\mathbb V_m = (V_m^t, V_m^b)$ in spinor (layer) space for different potential amplitudes on the top and bottom layers, allowing the model to mimic the ``moir\'e-distant" ($\left|V_b\right|\gg\left|V_t\right|$) and ``moir\'e-proximate" ($\left|V_t\right|\gg\left|V_b\right|$) geometries in rhombohedral graphene. When $V_m>0$, each superlattice cell contains two degenerate potential minima, forming a honeycomb lattice; when $V_m<0$, each superlattice cell has a single minimum, corresponding to a triangular lattice.

The moir\'e potential defines a mini-Brillouin zone (mBZ) and folds the basin band into three minibands. We measure the carrier density by the moir\'e filling factor $\nu$, defined so that $\nu=1$ corresponds to one carrier per moir\'e unit cell. Figs.~\ref{fig:bands}(c) and (d) show the moir\'e minibands for $\mathbb V_m = (0,8)$ and $(8,0)$, representing the moir\'e-distant and moir\'e-proximate cases. In the former, the moir\'e potential on the bottom layer separates the second and third minibands but leaves the lowest and second minibands overlapping. In the latter, the lower two bands are pushed apart near the mBZ boundary, although their crossings at $\kappa$ and $\kappa'$ remain protected by the honeycomb geometry, while the second and third bands stay nearly degenerate at $\gamma$. In both cases, moir\'e modulation does not open a gap between the lowest two minibands, so any insulating states at $\nu \leq 1$ must be driven by interactions.

{\it Integer Chern insulators.---}Now we consider Coulomb interactions $V_q = 2\pi U/q$ between electrons (the $q=0$ component is set to zero by a neutralizing background), where $U$ sets the interaction energy scale, and perform ED calculations with the system Hamiltonian projected onto the basin band. Details on the system Hamiltonian and band projection are presented in the Supplemental Material (SM)~\cite{SM}. Fig.~\ref{fig:integer_Chern}(a) shows the many-body energy spectrum at moir\'e filling factor $\nu=1$ and $\mathbb V_m = (0,8), U=20$, where a nondegenerate ground state is gapped from the higher-energy states. In the SM~\cite{SM} we present calculations of the spectral flow under flux insertion and the many-body Berry curvature~\cite{niu1985quantized}, establishing that this ground state is a Chern insulator with many-body Chern number $C=1$. Similar results are found for $\mathbb V_m=(8,0)$ [Fig.~\ref{fig:integer_Chern}(c)].

Figs.~\ref{fig:integer_Chern}(b) and (d) show the occupation of single-particle basin-band Bloch states in the many-body ground states in (a) and (c). We find that when the moir\'e potential is applied to the bottom layer, the occupation is nearly uniform across the mesh, whereas a moir\'e potential on the top layer concentrates occupation near the band bottom. This shows that the moir\'e-proximate Chern insulator is dominated by low-energy states at the band bottom, while the moir\'e-distant one is built from a mixture of band-bottom and dispersive states, occurring because the bottom-layer moir\'e potential pulls some of the dispersive states down into energetic overlap with the flat region [Fig.~\ref{fig:bands}(c)]. %Gapping these overlapping bands requires stronger interactions than gapping the point crossings left by the top-layer potential.
%While both Chern insulators are induced by interactions, the different Bloch state occupations suggest that the moir\'e-distant Chern insulator is the more strongly correlated state of the two.

\begin{figure}
    \centering
    \includegraphics[width=\linewidth]{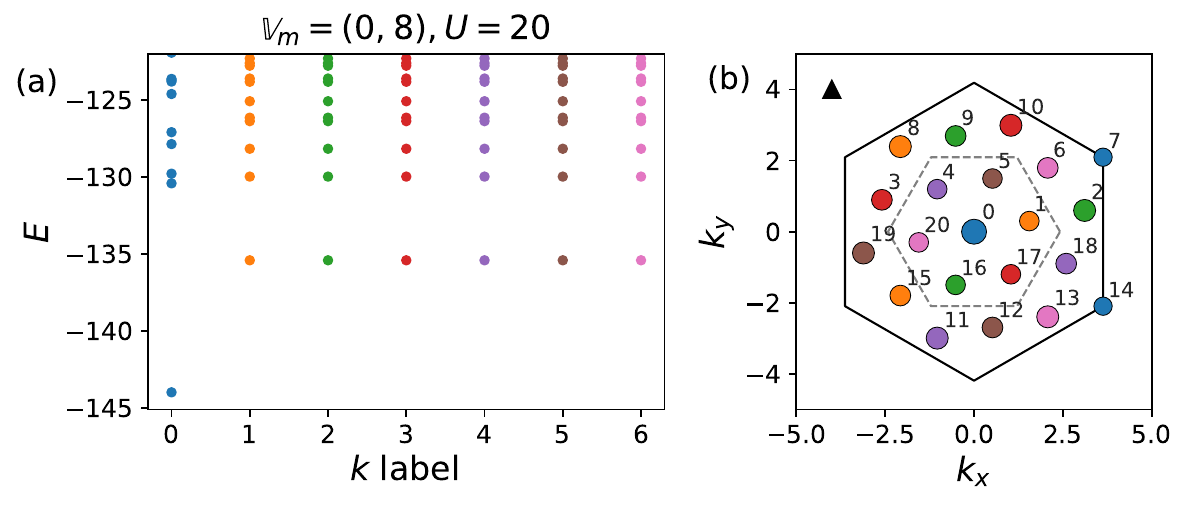}
    \includegraphics[width=\linewidth]{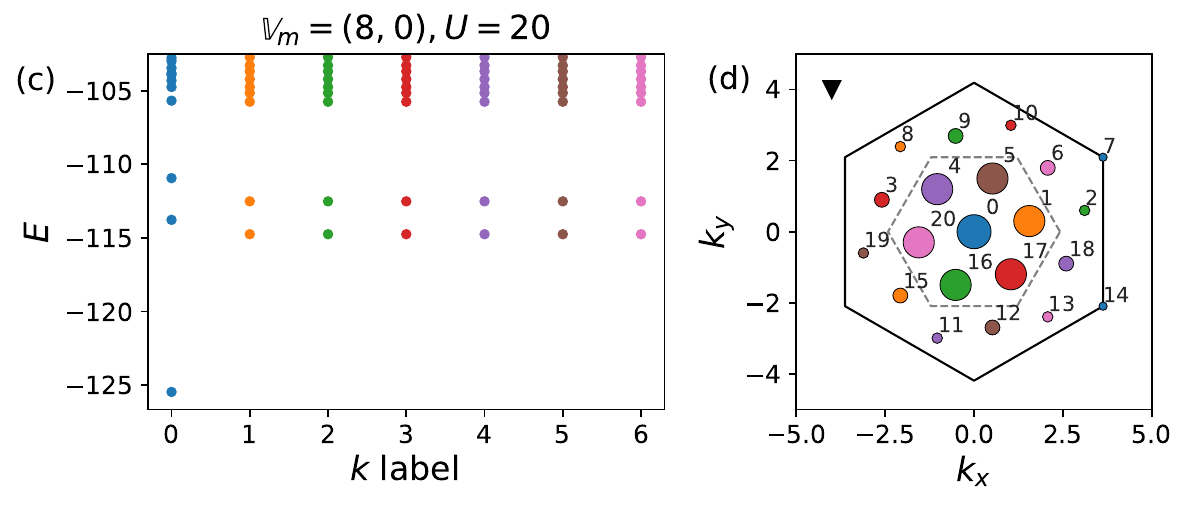}
    \includegraphics[width=\linewidth]{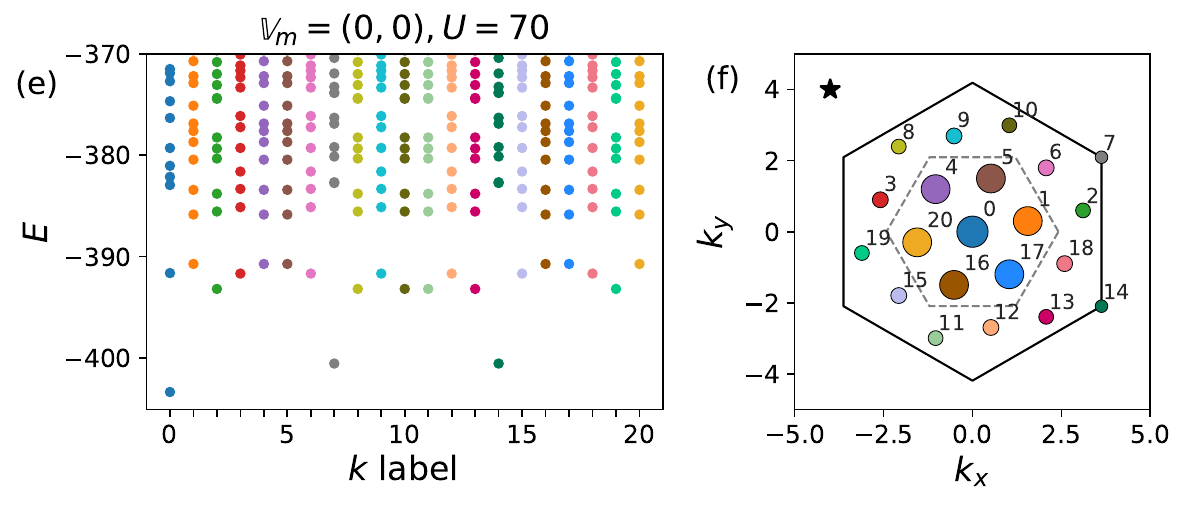}
    \caption{Integer Chern insulators at moir\'e filling $\nu=1$.   (a) Many-body energy spectrum for the remote potential case $\mathbb V_m = (0,8), U=20$ grouped by total-momentum sector, for $N_e=7$ electrons on the $N_s=21$ momentum mesh shown in (b); results for other system sizes are presented in the SM~\cite{SM}. (b) Sketch of momentum space with the large Brillouin zone (BZ) defined by the strong period potential $V_0$ shown as solid lines and the mini-BZ (mBZ) defined by moir\'e potential $V_m$ as dashed lines, momenta labelled by integers and colored to match the sectors in (a), and occupation of the single-particle Bloch states in the ground state of (a) shown as filled circles, with the area of each $\bm k$ point proportional to $\braket{n_{\bm k}}$. (c), (d) Same as (a), (b) but for the proximal potential case $\mathbb V_m = (8,0)$. (e), (f) Same as (a), (b) but for $V_m=0, U=70$, and the occupations are averaged over the three near-degenerate ground states of the crystal phase.}
    \label{fig:integer_Chern}
\end{figure}

Strong interactions can stabilize the Chern insulator ground state even in the absence of $V_m$. Fig.~\ref{fig:integer_Chern}(e) shows the many-body energy spectrum at $V_m=0, U=70$, where three near-degenerate ground states are separated from the excited states by a clear gap. The structure factor, spectral flow, and many-body Chern number calculations (see SM~\cite{SM}) show that the three-fold degeneracy arises from the spontaneous breaking of the translation symmetry of the parent lattice down to that of the $\sqrt{3}\times\sqrt{3}$ superlattice, and that the total Chern number of these three states is $C_{\rm tot}=3$, so each ground state is a commensurate anomalous Hall crystal~\cite{zeng2024sublattice, tan2024parent, soejima2024anomalous, dong2024stability, sheng2024quantum} with Chern number $C=1$.

\begin{figure}
    \centering
   \includegraphics[width=\linewidth]{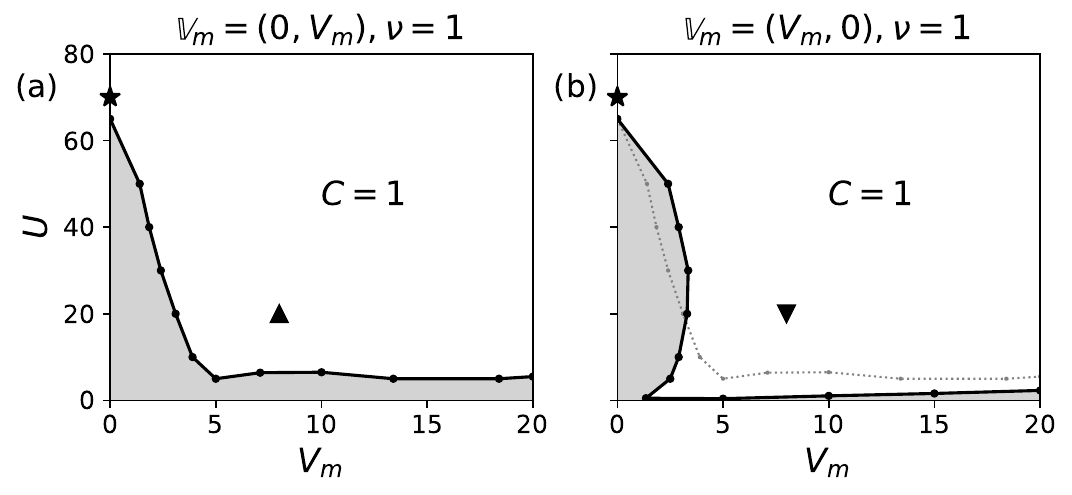}
    \includegraphics[width=\linewidth]{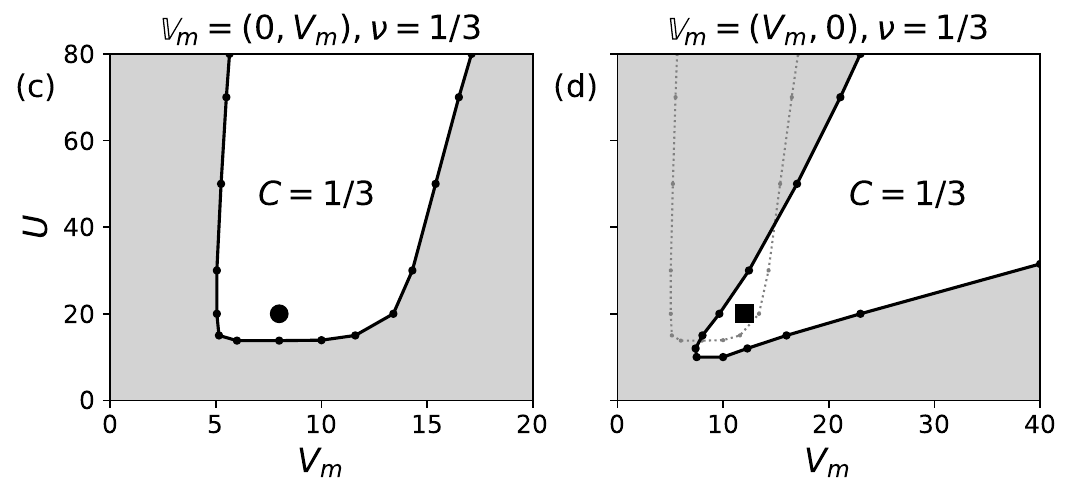}
    \caption{Phase diagrams at filling $\nu=1$ (a, b) and $\nu=1/3$ (c, d). Note the difference of $V_m$ scale in (d). The gray and white regions represent metallic states and insulating states with Hall conductance $\sigma_{xy}=-Ce^2/h$. The moir\'e-distant phase boundaries in (a) and (c) are plotted in the moir\'e-proximate phase diagrams (b) and (d) as gray dotted curves for comparison. The star and triangle symbols in (a) and (b) mark the parameters used in Fig.~\ref{fig:integer_Chern}, and the circle and square symbols in (c) and (d) mark the parameters used in Fig.~\ref{fig:FCI}.}
    \label{fig:phasediagram}
\end{figure}

We now consider the metal-insulator phase diagrams obtained by comparing the energy span of the ground state manifold under flux insertion against its gap to excited states; an insulating phase is defined by the latter exceeding the former. Figs.~\ref{fig:phasediagram}(a) and (b) show the phase diagrams in the $(V_m,U)$ plane for the remote potential (a) and proximal potential (b) cases. The two phase diagrams are similar, with the $C=1$ Chern insulator occupying a broad region of phase space at large $V_m$ or large $U$ and bounded by a metallic phase when at least one of $V_m$ and $U$ is small. Some differences are apparent.  The critical $U$ required to induce the Chern insulator is much smaller in the proximal potential case ($U_c\approx 1$ for $V_m=10$)  than in the remote case ($U_c\approx 5$ for $V_m\gtrsim 5$), while for $U\gtrsim 20$ the critical $V_m$ required is somewhat less in the remote potential case. In other words, a moir\'e potential on the remote layer is more effective at stabilizing strongly correlated insulating states although less effective at gapping the system in the weakly correlated regime. %In both cases the $C=1$ Chern insulator occupies a broad region, bounded by a metallic phase at small $V_m$ or $U$. The metal-insulator boundary is obtained by comparing the energy span of the ground state manifold under flux insertion against its gap to excited states; an insulating phase is defined by the latter exceeding the former. At $V_m\gtrsim 5$, gapping the system requires a finite $U\simeq 5$ when $V_m$ acts on the bottom layer, whereas a much smaller $U\lesssim 1$ suffices when $V_m$ acts on the top layer. The ordering reverses at strong coupling: for $U\gtrsim 20$ and below the crystallization transition at $U\simeq 65$, a finite $V_m$ is needed in both cases, but the critical $V_m$ is smaller when it acts on the bottom layer. The contrasting behavior at strong and weak coupling shows that a moir\'e potential on the bottom layer is more effective at stabilizing strongly correlated insulating states although less effective at gapping the system in the weakly correlated regime.

\begin{figure}
    \centering
    \includegraphics[width=\linewidth]{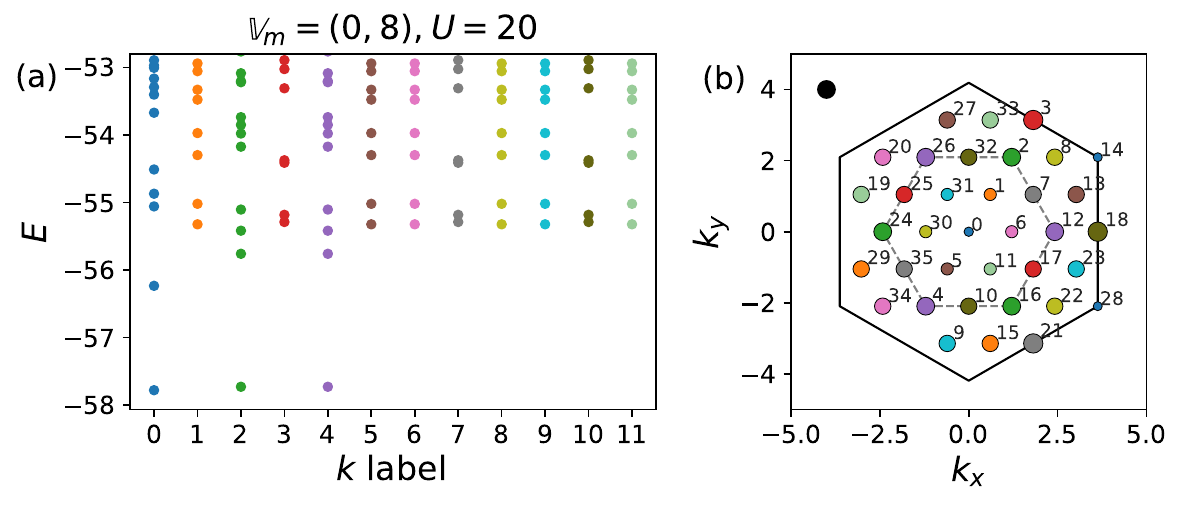}
    \includegraphics[width=\linewidth]{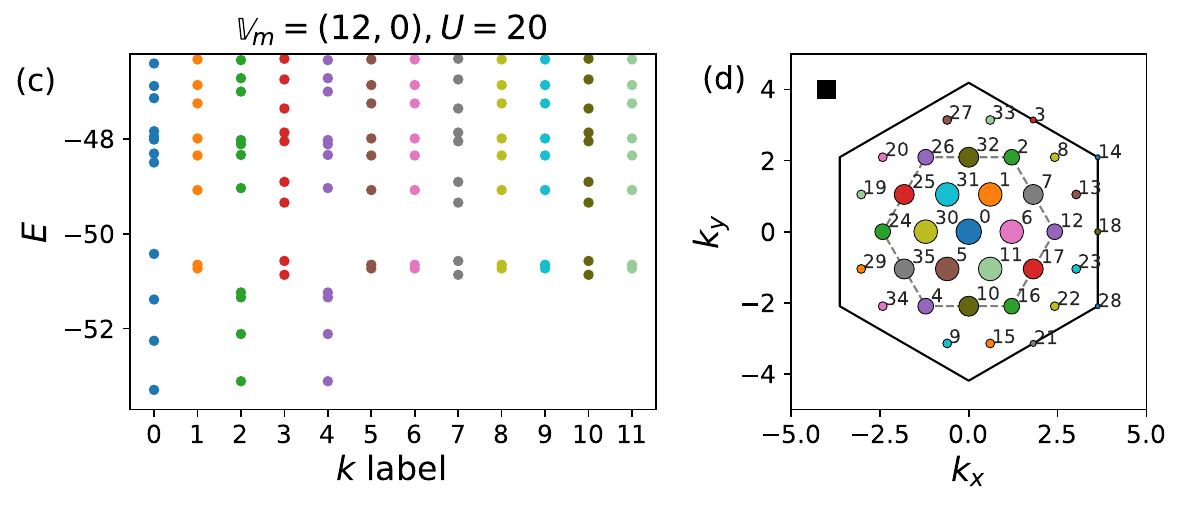}
    \caption{FCIs at moir\'e filling $\nu=1/3$ with different momenta distinguished both by an integer label and by color. (a) Many-body energy spectrum for $\mathbb V_m = (0,8), U=20$, for $N_e=4$ electrons on the $N_s=36$ momentum mesh shown in (b); results for other system sizes are presented in the SM~\cite{SM}. (b) Occupation of the single-particle Bloch state on the basin band $\braket{n_{\bm k}}$ averaged over the three near-degenerate ground states in (a), with the area of each $\bm k$ point proportional to $\braket{n_{\bm k}}$. (c), (d) Same as (a), (b) but for $\mathbb V_m = (12,0)$. (e), (f) Phase diagrams in $(V_m,U)$ plane for $V_m$ applied to the bottom and top layer, respectively. The gray and white regions represent metallic states and FCIs with $\sigma_{xy} = -e^2/3h$. The circle and square symbols mark the parameters used in (a)-(d).}
    \label{fig:FCI}
\end{figure}

{\it Fractional Chern insulators.---}We now turn to moir\'e filling $\nu=1/3$, i.e., filling $1/9$ of the parent basin band. Fig.~\ref{fig:FCI}(a) shows the many-body energy spectrum for $\mathbb V_m = (0,8), U=20$, where three near-degenerate ground states are separated from the excited states by a clear gap. The three-fold degeneracy, together with their total many-body Chern number $C_{\rm tot}=1$, identifies the ground state as an FCI with Hall conductance $\sigma_{xy} = -e^2/3h$. Additional diagnostic evidence, including the structure factor, spectral flow, and particle entanglement spectrum, is presented in the SM~\cite{SM}. Similar results are found when a larger $V_m$ acts on the top layer [Fig.~\ref{fig:FCI}(c)]. Figs.~\ref{fig:FCI}(b) and (d) show the corresponding Bloch state occupations averaged over the three ground states. Comparison of Figs.~\ref{fig:FCI}(a) and (c) shows that even though the moir\'e potential is stronger in the proximal potential case, the energy gap separating the ground state and excited states is smaller than the gap in the remote potential case, indicating that a moir\'e potential applied to the remote layer is more effective in stabilizing the FCI state.

Comparison of the ground state momentum distributions shown in Figs.~\ref{fig:FCI}(b,d) to the moir\'e bands in Figs.~\ref{fig:bands}(c,d) show that occupation is enhanced at momenta where the basin band is polarized to the layer to which $V_m$ is applied: inside the mBZ for the proximal potential case and outside when the potential is applied to the remote layer. The large weight at momenta outside the mBZ in the remote potential case indicates the importance of band mixing in this situation.

Figs.~\ref{fig:phasediagram}(c) and (d) show the $\nu=1/3$ phase diagrams in the $(V_m,U)$ plane. In both moir\'e-distant and moir\'e-proximate geometries, the FCI phase occupies a finite window in $V_m$ above a critical $U$. In both cases the FCI region never reaches $V_m=0$: in the absence of the moir\'e potential we do not find the fractional anomalous Hall crystal phase in our model at any interaction strength we access. For $V_m$ on the bottom layer, the boundaries of the FCI window are nearly independent of $U$, lying at $V_m\approx 5$ and $15$. We note that this window approximately coincides with the range of $V_m$ in which the lower two moir\'e minibands overlap (see SM~\cite{SM}). For $V_m$ on the top layer, both the lower and upper boundaries lie at considerably larger $V_m$ (note the different scale of the horizontal axis) and expand rapidly outward with increasing $U$.

{\it Discussion.---}Our minimal model provides insight into the circumstances under which integer and fractional Chern insulating states can develop. In the absence of a moir\'e potential, we find that very strong interactions can cause spontaneous translational symmetry breaking, producing an anomalous Hall crystal that is an integer Chern insulator. However, spontaneous translational symmetry breaking leading to a fractional anomalous Hall crystal is not found in our model, although it is not forbidden by fundamental principles. We find that adding a moir\'e superlattice potential stabilizes the integer Chern insulator state over a much wider range of interactions and plays an essential role in stabilizing fractional Chern insulator ground states. The sublattice structure of the potential is also important: for a triangular potential $V_m<0$ with a single minimum in each superlattice unit cell, we find (see SM~\cite{SM}) no FCIs at $\nu=1/3$, and the $C=1$ Chern insulators at $\nu=1$ survive only at strong interactions. This is consistent with our earlier finding~\cite{zeng2024sublattice, zeng2024gate} that topological states, in both single-particle and strong-interaction regimes, are favored by honeycomb lattices due to their sublattice degree of freedom.

Our model reproduces some of the key features of the experimental phenomenology observed in rhombohedral graphene; in particular the heretofore mysterious observation that applying the potential to the remote layer is more effective at stabilizing the FCI state than is applying the potential to the proximal layer. We find that rather than opening a single-particle gap to isolate the flat bottom of the basin band from its dispersive part, the bottom-layer moir\'e potential in our model stabilizes FCIs by lowering the energy of the dispersive states, leaving the lower two bands overlapping [Fig.~\ref{fig:bands}(c)]. This is most effective in the strong-interaction regime, where the FCI ground states are built from both low-energy states at the band bottom and high-energy states near the BZ boundary, so that moir\'e modulation makes them more favorable against competing Fermi liquids dominated by states at the band bottom. We expect a similar mechanism in rhombohedral graphene: while electrons at the conduction band bottom are strongly layer-polarized and barely feel a moir\'e potential acting on the opposite side of the sample, those in the dispersive region are spread more evenly across layers and gain more energy from the moir\'e modulation. The robustness of FCIs in the moir\'e-distant geometry thus indicates that these states are intrinsically multiband, drawing on single-particle states across a wide energy range rather than on the band bottom alone. This is consistent with the composite-fermion description developed in Ref.~\cite{ding2026composite}, where the lowest composite-fermion Landau level is delocalized across layers, allowing a remote moir\'e potential to open a sizable gap.
%While a moir\'e potential on the bottom layer does not separate the lowest two minibands at the single-particle level, strong interactions mix the top-layer-polarized states at the band bottom with the bottom-layer-polarized states in the dispersive region, and the moir\'e potential stabilizes the resulting correlated states by lowering the energy of these high-energy components. 
%Since both the integer and fractional quantum anomalous Hall states are correlated insulators built from a mixture of low- and high-energy electrons, a moir\'e potential acting on the high-energy components is more effective at stabilizing these states against competing Fermi liquids dominated by states at the band bottom.

Our results identify a new route to FCIs that does not require an isolated topological flat band. The ingredients are a dispersive band with a flat bottom and strong Berry curvature, together with a superlattice modulation that folds the band into the flat region but need not open a single-particle gap. Without fine-tuning the Bloch spinors toward an ``ideal" quantum geometry~\cite{roy2014band, jackson2015geometric, claassen2015position, wang2021exact, ledwith2023vortexability}, the FCI phase occupies a broad window in both the moir\'e and interaction strengths. In contrast to the conventional paradigm in which FCIs emerge in a narrow band whose quantum geometry approaches that of the lowest Landau level, our findings establish a distinct class of FCIs stabilized by the interplay of band dispersion, quantum geometry, interactions, and superlattice modulation.

Our model provides a controlled numerical platform for testing competing theories of FCIs in rhombohedral graphene since the basin band is compact and energetically isolated while retaining the essential features of the microscopic problem. Candidate ground states, including those from Hartree-Fock band-projected ED~\cite{dong2024anomalous, dong2024theory, zhou2024fractional, guo2024fractional, huang2024self, huang2025fractional, yu2025moire, li2025multi, regnault2026moire} and from composite-fermion constructions without band projection~\cite{ding2026composite}, can be compared directly with numerically exact results through wave function overlaps and energies. A central question is whether the moir\'e potential merely pins a strongly correlated state that already exists as a low-lying excited state in the $V_m=0$ spectrum, or whether it enters more intrinsically by mixing many $V_m=0$ eigenstates over a wide energy range. Distinguishing these possibilities in a controlled setting is a natural next step, with direct implications for how the FCIs observed in rhombohedral graphene should be understood. Our model also provides a natural starting point for wider investigations, for example into the interplay between FCI, electron crystal, conventional metallic, and superconducting states.

{\it Acknowledgments.---}We thank Nicol\'as Morales-Dur\'an, Jennifer Cano, and Leonid Glazman for helpful discussions.
This work was supported by the NSF MRSEC program through Columbia in the Center for Advanced Electronic Materials (DMR-2614607).
Work at Yale was supported by the Air Force Office of Scientific Research (AFOSR) under Award No. FA95502510287  and by NSF Grant No. DMR-2410182.
This work was performed in part at the Aspen Center for Physics, which is supported by National Science Foundation grant PHY-2210452.
The Flatiron Institute is a division of the Simons Foundation.

\bibliography{ref}

% ------------------------------------------------------------------------ %
% ------------------------------------------------------------------------ %
% ------------------------------------------------------------------------ %

\onecolumngrid
\newpage
\makeatletter 

\begin{center}
\textbf{\large Supplemental Material:\\[0.5em] \@title} \\[10pt]
% authors \\
% \textit{Institution}
\end{center}
\vspace{20pt}

\setcounter{figure}{0}
\setcounter{section}{0}
\setcounter{equation}{0}

\renewcommand{\thefigure}{S\@arabic\c@figure}
\renewcommand{\theequation}{S\@arabic\c@equation}
\renewcommand{\thetable}{S\@arabic\c@table}
\renewcommand{\theHfigure}{suppfig.\arabic{figure}}
\renewcommand{\theHequation}{suppeq.\arabic{equation}}
\makeatother

%\twocolumngrid
%\appendix
%\newpage

% ------------------------------------------------------------------------ %
% ------------------------------------------------------------------------ %
% ------------------------------------------------------------------------ %

\section{Model and band-projected ED}
\label{sec:SM_model}

\subsection{Single-particle Hamiltonian}

The model consists of spinless fermions in two dimensions carrying a two-component internal spinor
\begin{equation}
\ket{u_{\bm p}} =
\begin{pmatrix} \cos\tfrac{\theta_p}{2} \\[2pt]
e^{-iN\phi_{\bm p}}\sin\tfrac{\theta_p}{2}\end{pmatrix},
\qquad
\theta_p = \pi\tanh\!\left[(\lambda p/\pi)^2\right],
\qquad
\phi_{\bm p} = \arg(p_x + ip_y),
\label{eq:SM_spinor}
\end{equation}
so that $\ket{u}$ points to the north pole at $\bm p = 0$, rotates smoothly
to the south pole as $p\to\infty$, and winds $N$ times in phase. The
associated Berry curvature is concentrated near $\bm p=0$ and carries
total flux $2\pi N$ through the plane. The kinetic energy is
\begin{equation}
\epsilon_{\bm p} = \frac{p^4}{4} - \beta p^2 .
\end{equation}

A periodic potential couples plane waves differing by the first-shell
reciprocal lattice vectors of a triangular lattice with
$\bm G_1 = (4\pi/\sqrt3,0)$, lattice constant $a=1$, and $C_3$ partners
$\bm G_2, \bm G_3$ satisfying $\bm G_1 + \bm G_2 + \bm G_3 = 0$. It
acts only on the second spinor component,
\begin{equation}
\bra{\bm p + \bm G} H_{\rm pot} \ket{\bm p}
= \braket{u_{\bm p + \bm G} | \mathcal V_0 | u_{\bm p}},
\qquad
\mathcal V_0 = \mathrm{diag}(0, V_0),
\qquad i = 1,2,3,
\end{equation}
with the matrix elements at $-\bm G_i$ fixed by hermiticity,
$\bra{\bm p - \bm G_i}H_{\rm SL}\ket{\bm p} = u(\bm p - \bm
G_i)^\dagger \mathcal V_0^* u(\bm p)$. In general
$V_0 = |V_0|e^{i\varphi_0}$ is complex; the phase $\varphi_0$ fixes the
registry of the potential relative to the lattice, and we take
$\varphi_0 = \pi$ (i.e.\ $V_0 < 0$) throughout.

On a discrete mesh $\{\bm k\}$ truncated to a disc $|\bm G|\le R_G$,
the Bloch Hamiltonian is
\begin{equation}
h_{\bm G'\bm G}(\bm k) =
\left(\frac{|\bm k+\bm G|^4}{4} - \beta|\bm k+\bm G|^2\right)
\delta_{\bm G'\bm G}
+ \braket{u_{\bm k+\bm G'} | \mathcal V_0^{(\bm G'-\bm G)} | u_{\bm k+\bm G}},
\label{eq:SM_bloch}
\end{equation}
where $\mathcal V_0^{(\bm G'-\bm G)} = \mathrm{diag}(0,V_0)$ for
$\bm G'-\bm G \in \{\bm G_1,\bm G_2,\bm G_3\}$,
$\mathrm{diag}(0,V_0^*)$ for the negatives, and zero otherwise. Its
eigenvectors $z_{\bm G}(\bm k)$ define the basin band and satisfy $z_{\bm G}(\bm k+\bm G') = z_{\bm G+\bm G'}(\bm k)$.

The moir\'e potential has the same structure on the
$\sqrt3\times\sqrt3$ shell. Its reciprocal vectors are
$\bm g_1 = (0, 4\pi/3)$ and its $C_3$ partners $\bm g_{2,3}$, again with
$\bm g_1 + \bm g_2 + \bm g_3 = 0$ and $|\bm g_i| = |\bm G_i|/\sqrt3$:
\begin{equation}
\bra{\bm p + \bm g_i} H_m \ket{\bm p}
= \braket{u_{\bm p + \bm g_i} | \mathcal V_m | u_{\bm p}},
\qquad
\mathcal V_m = \mathrm{diag}(V_m^t, V_m^b),
\label{eq:SM_moire_def}
\end{equation}
with $\bm g \to -\bm g$ again fixed by hermiticity. In real space the
corresponding scalar potential is
\begin{equation}
V_m(\bm r) = 2|V_m| \sum_{i=1}^{3}
\cos(\bm g_i\cdot\bm r + \varphi_m),
\end{equation}
so that the phase $\varphi_m$ controls the registry: $\varphi_m = 0$
places two degenerate minima per superlattice cell, forming a honeycomb
lattice, while $\varphi_m = \pi$ places a single minimum per cell,
forming a triangular lattice. Intermediate $\varphi_m$ breaks the
equivalence of the two honeycomb sublattices. Throughout this work we
restrict to the two symmetric values $\varphi_m = 0$ and $\pi$, which
we refer to as $V_m > 0$ and $V_m < 0$.

\begin{figure}
    \centering
    \includegraphics[width=\linewidth]{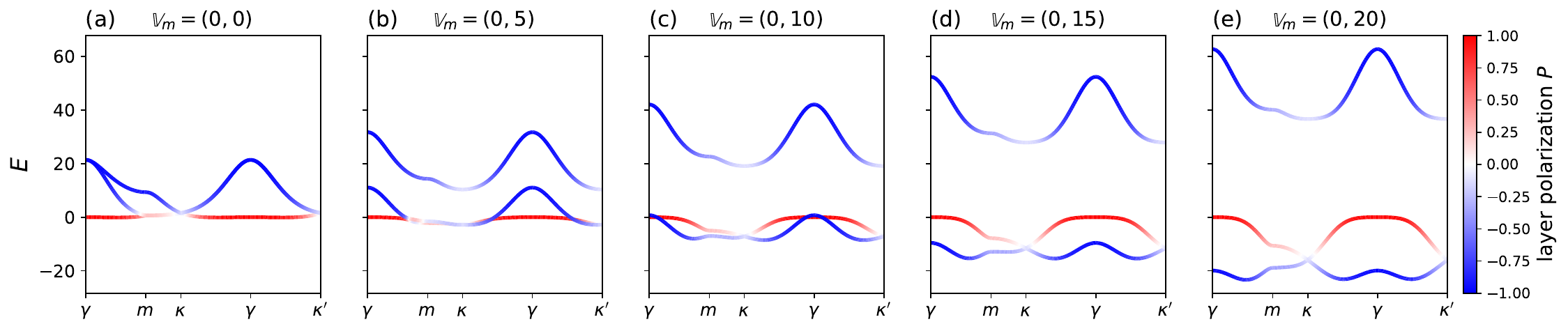}
    \caption{Single-particle band structure for moir\'e potential $V_m = 0,5,10,15,20$ applied to the {\it bottom} layer. Other parameters take the same values as in the main text: $N=3,\lambda=1,\beta=0.2,V_0=-25$.}
    \label{fig:bands_Vm_scan}
\end{figure}

\begin{figure}
    \centering
    \includegraphics[width=\linewidth]{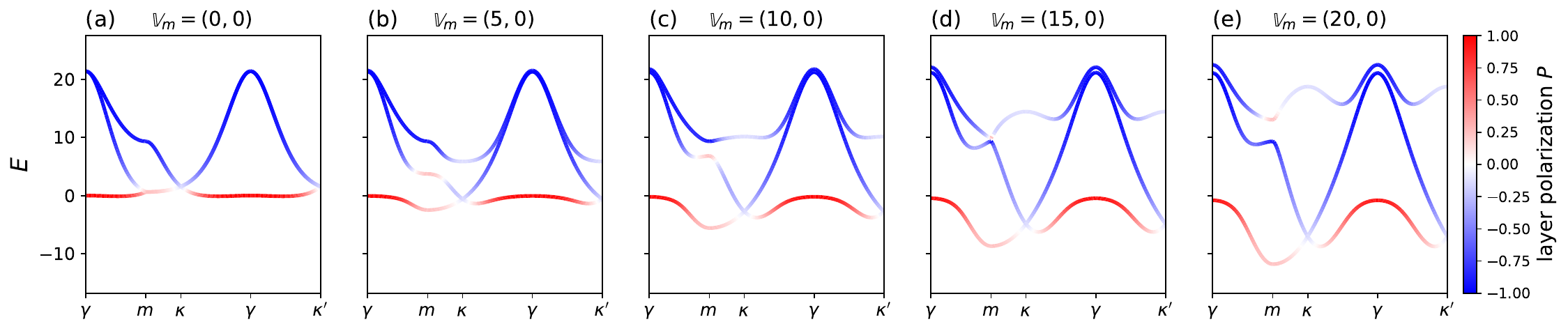}
    \caption{Single-particle band structure for moir\'e potential $V_m = 0,5,10,15,20$ applied to the {\it top} layer. Other parameters take the same values as in the main text: $N=3,\lambda=1,\beta=0.2,V_0=-25$.}
    \label{fig:bands_Vm_top_scan}
\end{figure}

Fig.~\ref{fig:bands_Vm_scan} shows the single-particle minibands for $V_m=0,5,10,15,20$ applied to the bottom layer. As $V_m$ increases, the upper two bands---polarized to the bottom layer---are pushed apart, while the lowest band stays largely unaffected except near the mBZ boundary. For $V_m\in (5,15)$, the lower two bands overlap across a large portion of the mBZ; remarkably, this interval aligns with the FCI stability window at filling $\nu=1/3$.

Fig.~\ref{fig:bands_Vm_top_scan} shows the single-particle bands for $V_m$ applied to the top layer. A large $V_m$ makes the lower two bands strongly dispersive near the Dirac points, while the gap between the second and third bands remains small because the Bloch states at the band top are strongly polarized to the bottom layer.

\subsection{Projected density operator and form factor}

The Coulomb interaction couples only to the total charge density, so
the density operator at momentum transfer $\bm q$ is
\begin{equation}
\rho_{\bm q} = \sum_{\bm p} \braket{u_{\bm p + \bm q} | u_{\bm p}}\,
c^\dagger_{\bm p + \bm q} c_{\bm p} .
\end{equation}
Projecting onto the basin band and writing $\bm p = \bm k + \bm G$,
\begin{equation}
\bar\rho_{\bm q} = \sum_{\bm k} \Lambda_{\bm q}(\bm k)\,
\gamma^\dagger_{[\bm k + \bm q]}\gamma_{\bm k},
\qquad
\Lambda_{\bm q}(\bm k) = \sum_{\bm G}
z_{\bm G}^*(\bm k+\bm q) z_{\bm G}(\bm k) \braket{u_{\bm{k+q+G}}|u_{\bm{k+G}}}
\label{eq:SM_formfactor}
\end{equation}
where $[\bm k + \bm q]$ denotes the mesh representative of
$\bm k + \bm q$. The form factor satisfies
$\Lambda_{\bm 0}(\bm k) = 1$, $|\Lambda_{\bm q}(\bm k)| \le 1$, and
$\Lambda_{\bm q}(\bm k)^* = \Lambda_{-\bm q}([\bm k + \bm q])$.

\subsection{Projected interaction}

The Coulomb Hamiltonian is
\begin{equation}
H_{\rm int} = \frac{1}{2A}\sum_{\bm q \ne 0} V_q
: \rho_{\bm q}\rho_{-\bm q} : ,
\qquad V(q) = \frac{2\pi U}{q},
\end{equation}
normal-ordered with respect to the empty vacuum; the $\bm q = 0$ term
is cancelled by a neutralizing background. Here $A = \sqrt{3}N_s/2$ is the area of the system with $N_s$ unit cells. Since all four operators in
$:\rho\rho:$ act on states built from the lowest band, projection acts
independently on each leg, giving
\begin{equation}
\bar H_{\rm int} = \frac{1}{2A}\sum_{\bm q\ne 0} V(q)
\sum_{\bm k_1 \bm k_2}
\Lambda_{\bm q}(\bm k_1)\Lambda_{-\bm q}(\bm k_2)\,
\gamma^\dagger_{[\bm k_1 + \bm q]}\gamma^\dagger_{[\bm k_2 - \bm q]}
\gamma_{\bm k_2}\gamma_{\bm k_1} .
\end{equation}

Writing $\bm k_3 = [\bm k_1 + \bm q]$ and $\bm k_4 = [\bm k_2 - \bm
q]$, the momentum-conserving matrix element is
\begin{equation}
V^{\bm k_3\bm k_4}_{\bm k_2\bm k_1} = \frac{1}{2A}
{\sum_{\bm G}}'
V(|\bm k_3 - \bm k_1 + \bm G|)\,
\Lambda_{\bm k_3 - \bm k_1 + \bm G}(\bm k_1)\,
\Lambda_{-(\bm k_3 - \bm k_1 + \bm G)}(\bm k_2),
\label{eq:SM_matrixelement}
\end{equation}
where the prime excludes only the single term with $\bm k_3 - \bm k_1 +
\bm G = 0$; all remaining umklapp channels, including the Hartree terms
with $\bm k_3 = \bm k_1$ and $\bm G \ne 0$, are retained. The
antisymmetrized, pair-ordered form used in the code is
\begin{equation}
\bar H_{\rm int} = \sum_{\bm k_1 < \bm k_2}\sum_{\bm k_3 < \bm k_4}
2\left[V^{\bm k_3\bm k_4}_{\bm k_2\bm k_1}
- V^{\bm k_4\bm k_3}_{\bm k_2\bm k_1}\right]
\gamma^\dagger_{\bm k_3}\gamma^\dagger_{\bm k_4}
\gamma_{\bm k_2}\gamma_{\bm k_1} .
\end{equation}

\subsection{Moir\'e potential in the projected basis}
\label{sec:SM_moire}

Projected onto the basin band, the moir\'e potential is a one-body
operator that changes momentum by $\bm g \in \{\pm\bm g_1, \pm\bm g_2,
\pm\bm g_3\}$:
\begin{equation}
\bar H_m = \sum_{\bm k}\sum_{\bm g}
\mathcal M_{\bm g}(\bm k)\, \gamma^\dagger_{[\bm k+\bm g]}\gamma_{\bm k},
\quad
\mathcal M_{\bm g}(\bm k) = \sum_{\bm G}
z^*_{\bm G}(\bm k+\bm g) z_{\bm G}(\bm k)
\braket{u_{\bm k+\bm g+\bm G} | \mathcal V_m^{(\bm g)} | u_{\bm k+\bm G}},
\end{equation}
where $\mathcal V_m^{(\bm g)}$ carries $V_m$ for $\bm g \in \{\bm g_1,\bm
g_2,\bm g_3\}$ and $V_m^*$ for their negatives. This guarantees
$\mathcal M_{\bm g}(\bm k)^* = \mathcal M_{-\bm g}([\bm k+\bm g])$ and
hence $\bar H_m^\dagger = \bar H_m$.

\subsection{Simulation torus and diagonalization}

We work on a torus defined by an integer generator matrix
$M = \begin{pmatrix} N_1 & 0 \\ t & N_2\end{pmatrix}$,
% >> or the general 2x2 integer form if the code allows it; state which.
so that the allowed momenta are
\begin{equation}
\bm k(\bm n) = \left[(\bm n + \bm\phi/2\pi)\, M^{-1}\right]_1 \bm G_1
+ \left[(\bm n + \bm\phi/2\pi)\, M^{-1}\right]_2 \bm G_2 ,
\qquad \bm n \in \mathbb Z^2 ,
\end{equation}
with $N_s = |\det M|$ inequivalent points and boundary twists
$\bm\phi = (\phi_1,\phi_2)$ used for spectral flow and many-body Chern
numbers. Nonzero tilt $t$ gives clusters that are more isotropic at the same $N_s$ and, more
importantly, reaches system sizes unavailable to the untilted family.

Commensuration with the $\sqrt3\times\sqrt3$ superlattice requires the
mesh to be invariant under shifts by the moir\'e reciprocal vectors,
equivalently that it contain the three points $\Gamma$, $K$, and $K'$
of the parent Brillouin zone. For a
untilted mesh this requires $3|N_1$ and $3|N_2$; tilted clusters
satisfy it for additional $(N_1,N_2,t)$, including $N_s$ not divisible
by $9$.
The clusters used in this work are listed in
Table~\ref{tab:SM_clusters}.

\begin{table}[t]
\caption{\label{tab:SM_clusters}
Clusters used in this work.
$N_e$: electron number; $\nu$: moir\'e filling factor.}
\begin{tabular}{cccccc}
\hline\hline
$\nu$ & $N_s$ & $N_e$ & $(N_1,N_2,t)$  \\
\hline
$1$   & $21$ & $7$ & $(1,21,4)$ \\
$1$   & $24$ & $8$ & $(1,24,4)$ \\
$1$   & $27$ & $9$ & $(3,9,1)$ \\
$1/3$ & $36$ & $4$ & $(6,6,0)$ \\
$1/3$ & $45$ & $5$ & $(3,15,2)$ \\
$1/3$ & $54$ & $6$ & $(6,9,0)$ \\
\hline\hline
\end{tabular}
\end{table}

\section{Additional diagnostics}

\subsection{Integer Chern insulators}

\begin{figure}
    \centering
    \includegraphics[width=0.3\linewidth]{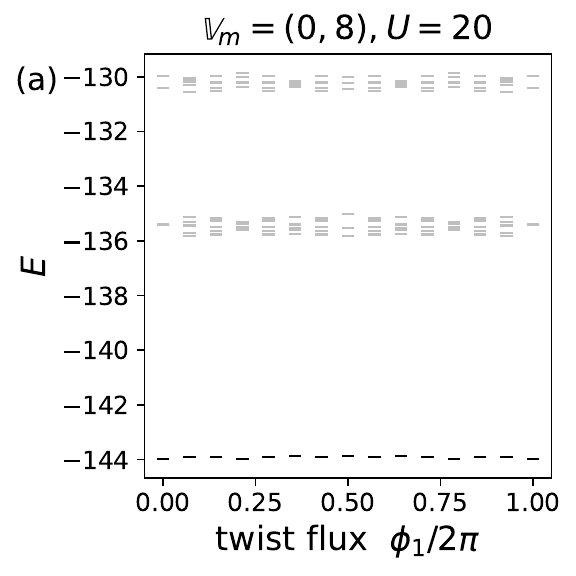}
    \includegraphics[width=0.3\linewidth]{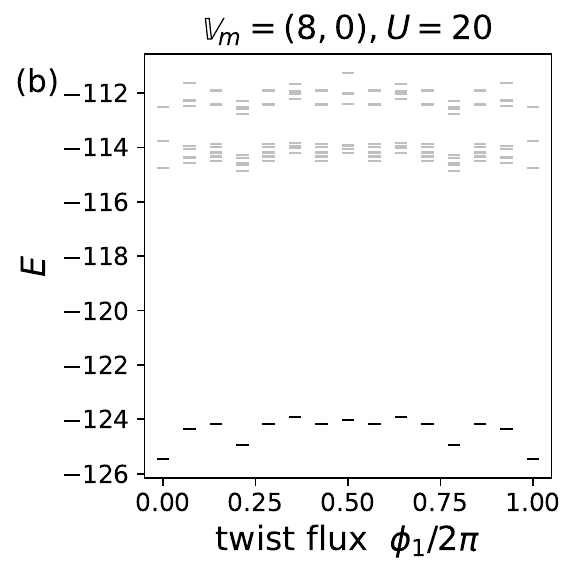}
    \includegraphics[width=0.3\linewidth]{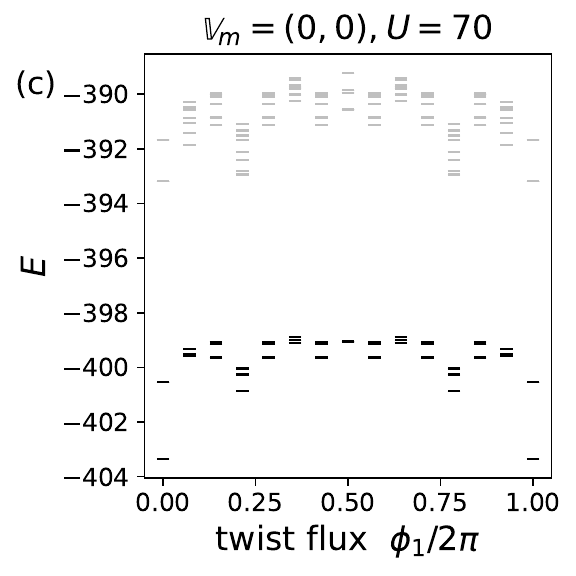}
    \caption{Spectral flow under flux insertion $\phi_1$ for $N_s=21, N_e=7$, and (a) $\mathbb V_m=(0,8), U=20$; (b) $\mathbb V_m=(8,0), U=20$; (c) $V_m=0, U=70$.}
    \label{fig:flow_int}
\end{figure}

Fig.~\ref{fig:flow_int} shows the evolution of the low-lying spectrum under
insertion of a boundary twist $\phi_1$. For $\mathbb V_m = (0,8)$ and
$(8,0)$ at $U=20$ the nondegenerate ground state remains separated from
the excited states by a gap of order 8-10
throughout the flux cycle, returning to itself after one flux quantum.
The absence of any level crossing with the excited spectrum confirms
incompressibility and guarantees that the many-body Chern number
extracted below is well defined.

At $V_m = 0$ and $U = 70$ [Fig.~\ref{fig:flow_int}(c)] the three
near-degenerate ground states permute but remain isolated from the rest of the
spectrum for all $\phi_1$, confirming the incompressibility of the ground state manifold. Since $U=70$ is close to the crystallization phase boundary ($U_c\simeq 65$), the gap to excited states is only slightly larger than the energy span of the ground state manifold.

\begin{figure}
    \centering
    \includegraphics[width=0.3\linewidth]{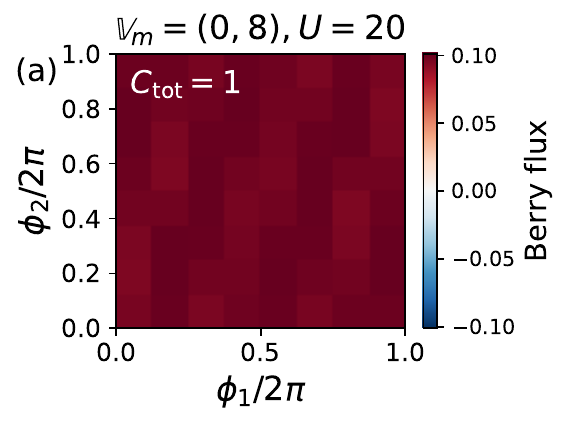}
    \includegraphics[width=0.3\linewidth]{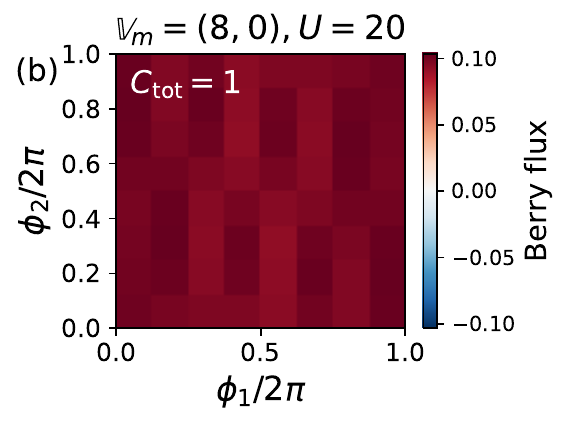}
    \includegraphics[width=0.3\linewidth]{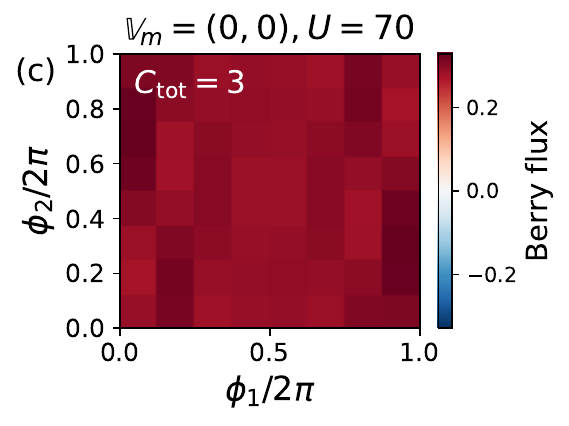}
    \caption{Many-body Berry curvature of the ground state manifold for $N_s=21, N_e=7$, and (a) $\mathbb V_m=(0,8), U=20$; (b) $\mathbb V_m=(8,0), U=20$; (c) $V_m=0, U=70$. The color scale shows the Berry flux in each plaquette on an $8\times 8$ $\phi$-grid. The total many-body Chern number is $C_{\rm tot}=1$ in (a), (b) and $C_{\rm tot}=3$ in (c).}
    \label{fig:berry_int}
\end{figure}

For a ground-state manifold of $D$ states $\{\ket{\Psi_n(\bm\phi)}\}$
that remains separated from the excited spectrum at every twist, we
compute the total Chern number of the manifold by the
plaquette method: On an $L\times L$ grid of twists $\bm\phi = (\phi_1,\phi_2)$ with spacing
$\delta\phi = 2\pi/L$, define the non-Abelian link variables
\begin{equation}
\left[U_\mu(\bm\phi)\right]_{mn} =
\braket{\Psi_m(\bm\phi)|\Psi_n(\bm\phi + \delta\phi\,\hat{\bm e}_\mu)},
\qquad \mu = 1,2 ,
\end{equation}
from which the Berry flux through the plaquette at $\bm\phi$ is
\begin{equation}
F(\bm\phi) = \arg \det\!\left[
U_1(\bm\phi)\,
U_2(\bm\phi + \delta\phi\,\hat{\bm e}_1)\,
U_1(\bm\phi + \delta\phi\,\hat{\bm e}_2)^{-1}\,
U_2(\bm\phi)^{-1}
\right] \in (-\pi,\pi] ,
\label{eq:SM_berryflux}
\end{equation}
and the total Chern number of the manifold is
\begin{equation}
C_{\rm tot} = \frac{1}{2\pi}\sum_{\bm\phi} F(\bm\phi) .
\label{eq:SM_chern}
\end{equation}
Fig.~\ref{fig:berry_int} shows the many-body Berry curvature of the
insulating states in Fig.~\ref{fig:flow_int}, computed on an $8\times 8$ grid of boundary twists
$(\phi_1,\phi_2)$. In all three cases, the Berry flux is positive and nearly uniform on every plaquette. Summing over plaquettes gives
$C_{\rm tot} = 1$ for both $\mathbb V_m = (0,8)$ and $(8,0)$ at
$U = 20$, and $C_{\rm tot} = 3$ for the three-fold manifold at
$V_m = 0$, $U = 70$. In the latter case the three states are related by
translation and therefore carry equal Chern numbers, so each is a
Chern insulator with $C = 1$; the corresponding Hall conductance is
$\sigma_{xy} = -e^2/h$ per state.

\begin{figure}
    \centering
    \includegraphics[width=0.3\linewidth]{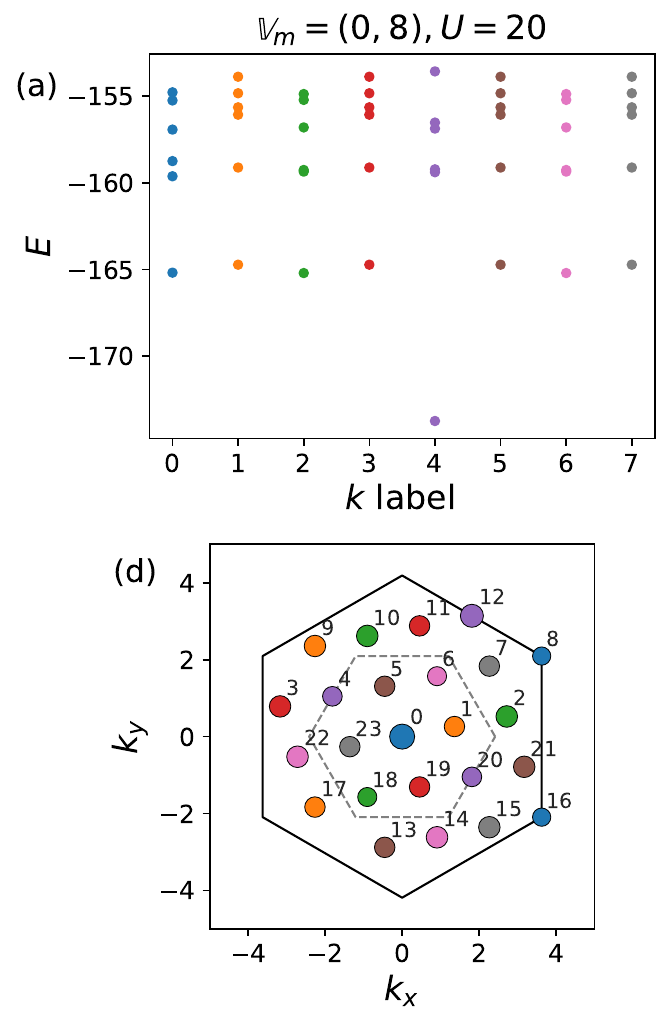}
    \includegraphics[width=0.3\linewidth]{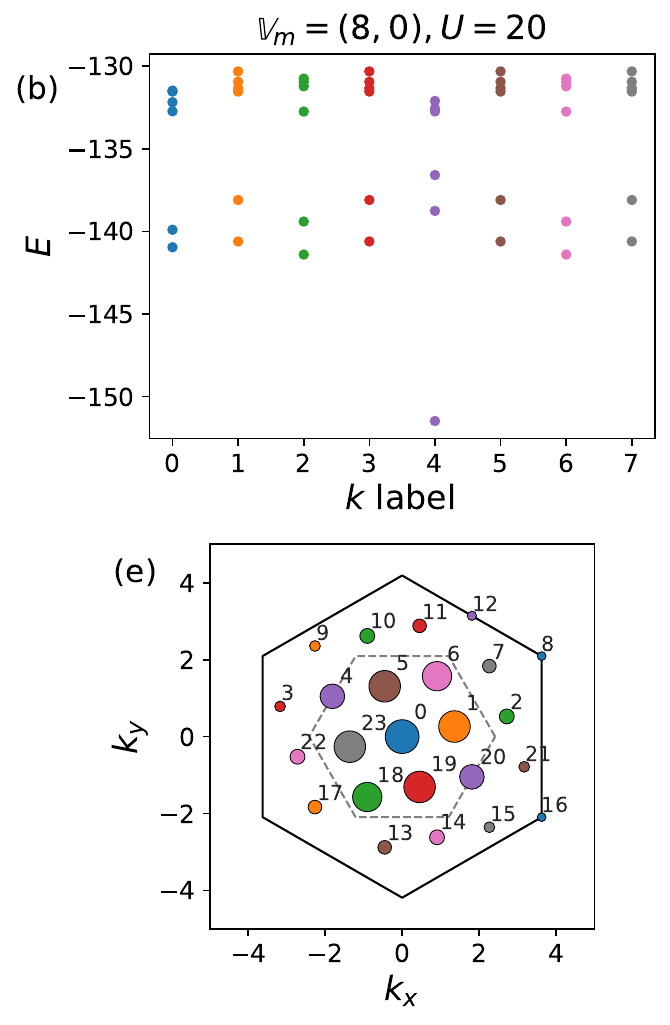}
    \includegraphics[width=0.3\linewidth]{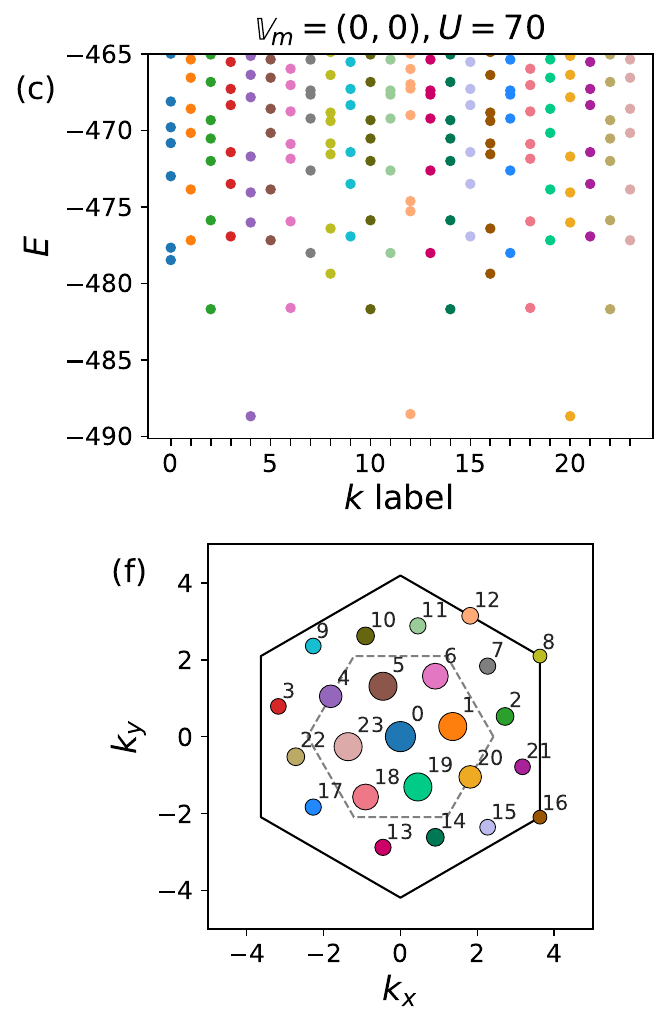}
    \caption{Many-body energy spectrum for $N_s=24, N_e=8$, and (a) $\mathbb V_m=(0,8), U=20$; (b) $\mathbb V_m=(8,0), U=20$; (c) $V_m=0, U=70$. (d)-(f) show the corresponding Bloch state occupations.}
    \label{fig:spectrum_int_24}
\end{figure}

\begin{figure}
    \centering
    \includegraphics[width=0.3\linewidth]{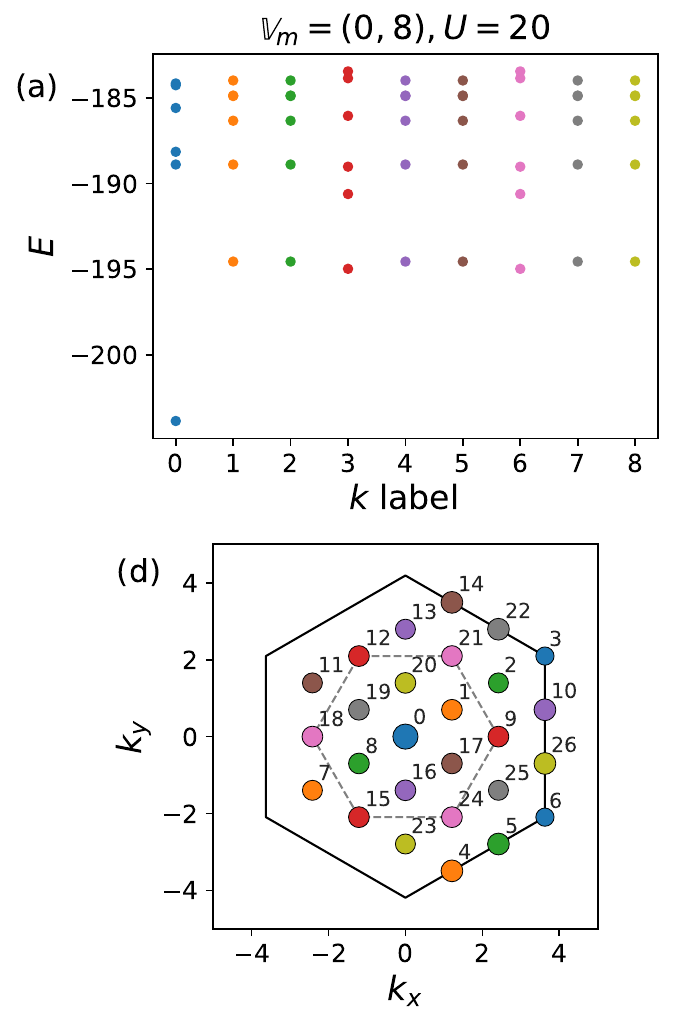}
    \includegraphics[width=0.3\linewidth]{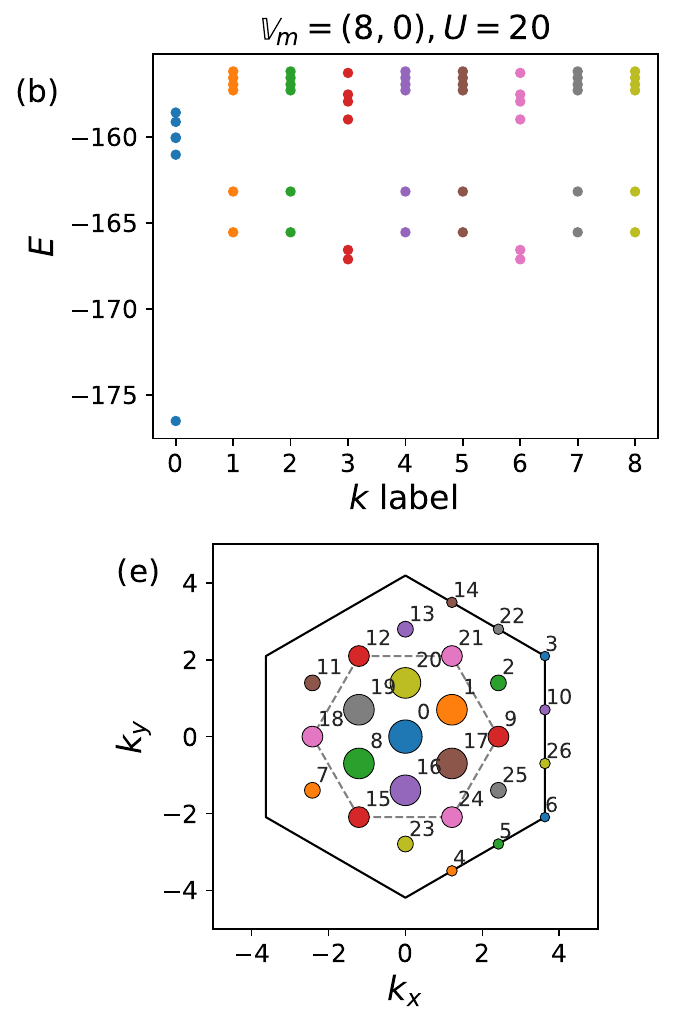}
    \includegraphics[width=0.3\linewidth]{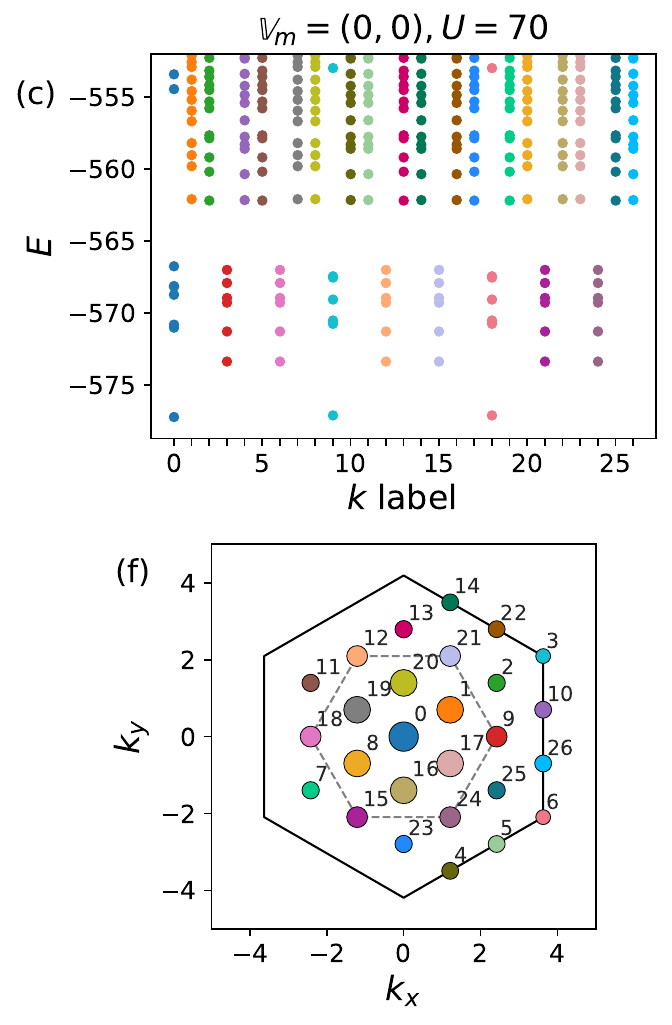}
    \caption{Many-body energy spectrum for $N_s=27, N_e=9$, and (a) $\mathbb V_m=(0,8), U=20$; (b) $\mathbb V_m=(8,0), U=20$; (c) $V_m=0, U=70$. (d)-(f) show the corresponding Bloch state occupations.}
    \label{fig:spectrum_int_27}
\end{figure}

Figs.~\ref{fig:spectrum_int_24} and~\ref{fig:spectrum_int_27} plot the ED results for
$N_s = 24$, $N_e = 8$ and $N_s = 27$, $N_e = 9$, both at moir\'e
filling $\nu = 1$. In both cases the moir\'e-stabilized states at
$\mathbb V_m = (0,8)$ and $(8,0)$ have a nondegenerate ground state
separated from the excited spectrum,
and at $V_m = 0$, $U = 70$ a threefold manifold appears at momenta
separated by $N_s/3$.
The Bloch-state occupations shown in the lower panels reproduce the
pattern discussed in the main text: the occupation is enhanced at the
momenta where the basin band is polarized to the layer on which $V_m$
acts. All results are quantitatively consistent across all system sizes, except that the many-body gap in Fig.~\ref{fig:spectrum_int_27}(c) for $N_s=27$ is substantially smaller than that for $N_s=21$ and $24$. As noted above, the parameter choice is close to the crystallization phase boundary, so the many-body gap is not robust and finite-size correction is relatively large. Nevertheless, for all three system sizes, the three-fold degenerate ground state manifold is clearly separated from the excited states.

\begin{figure}
    \centering
    \includegraphics[width=0.3\linewidth]{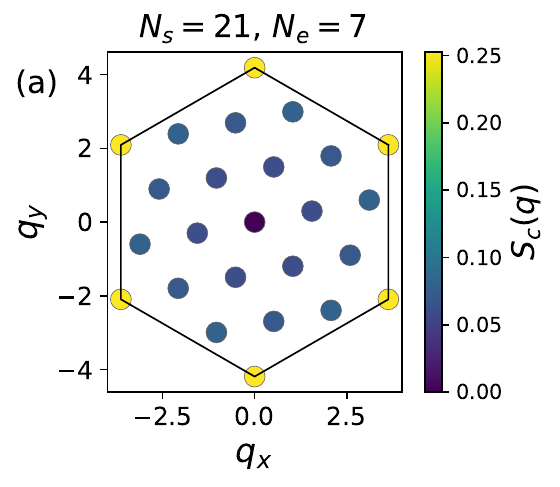}
    \includegraphics[width=0.3\linewidth]{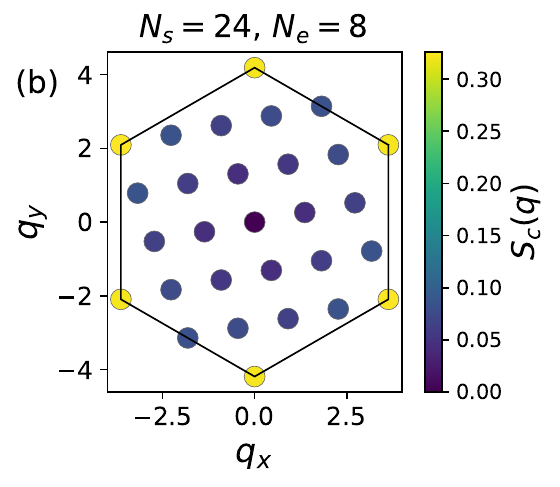}
    \includegraphics[width=0.3\linewidth]{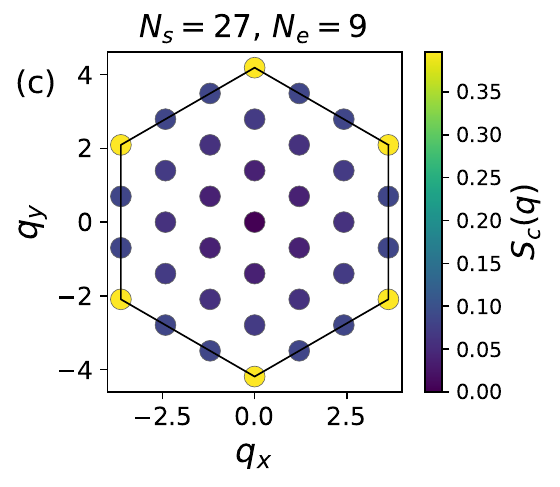}
    \caption{Structure factor $S_c(\bm q)$ averaged over the three ground states at $V_m=0, U=70$, for three system sizes: (a) $N_s=21, N_e=7$; (b) $N_s=24, N_e=8$; (c) $N_s=27, N_e=9$.}
    \label{fig:Sq_AHC}
\end{figure}

To establish that the three-fold degeneracy at $V_m = 0$ reflects
genuine translation symmetry breaking, we compute the connected
structure factor
\begin{equation}
S_c(\bm q) = \frac{\langle \rho_{\bm q}\rho_{-\bm q}\rangle
- |\langle \rho_{\bm q}\rangle|^2}{N_e},
\end{equation}
averaged over the three ground states. Figure~\ref{fig:Sq_AHC} shows
$S_c(\bm q)$ for $N_s = 21$, $24$, and $27$. In all three cases the
structure factor is sharply peaked at the corners of the
BZ, i.e. at $\pm\bm K = \pm(2\pi/\sqrt{3},2\pi/3)$, which are precisely the ordering wavevectors of a $\sqrt3\times\sqrt3$ charge modulation. The peak height grows systematically with system size:
$S_c(\pm\bm K) \simeq 0.252$, $0.323$, and $0.396$ for $N_e = 7$, $8$, and $9$.
Since $S_c$ is already normalized by $N_e$, a peak that grows linearly rather
than saturates indicates that $\langle|\rho_{\pm\bm K}|^2\rangle$
grows faster than the particle number, as expected for long-range
crystalline order rather than short-range correlations.
Together with $C_{\rm tot} = 3$, this identifies the $V_m = 0, U=70$ ground
states as commensurate anomalous Hall crystals.

\subsection{Fractional Chern insulators}

\begin{figure}
    \centering
    \includegraphics[width=0.3\linewidth]{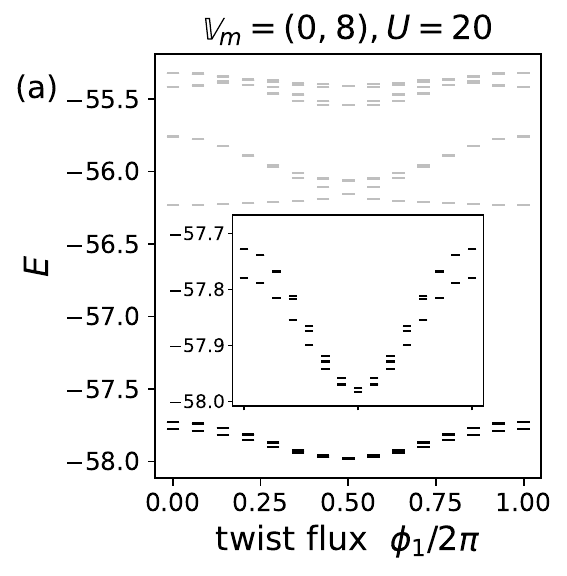}
    \includegraphics[width=0.3\linewidth]{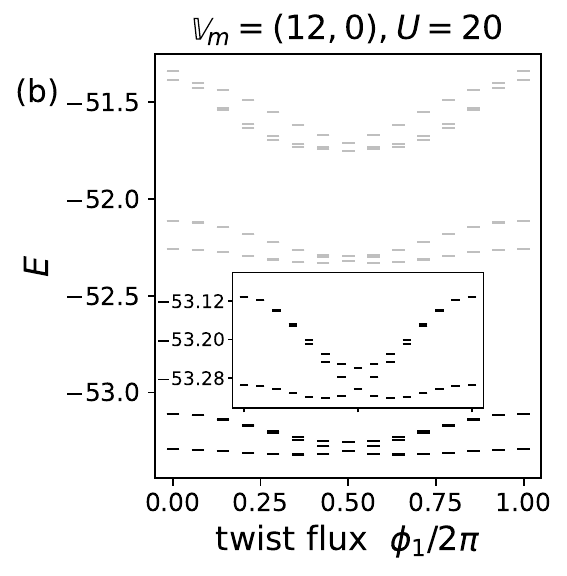}
    \caption{Spectral flow under flux insertion $\phi_1$ for $N_s=36, N_e=4$, and (a) $\mathbb V_m=(0,8), U=20$; (b) $\mathbb V_m=(12,0), U=20$. Insets: magnified view of the permutation of the lowest three states under flux insertion.}
    \label{fig:flow_FCI}
\end{figure}

Fig.~\ref{fig:flow_FCI} shows the low-lying spectrum under insertion of a
twist $\phi_1$ for $N_s = 36$, $N_e = 4$. In both
$\mathbb V_m = (0,8)$ and $(12,0)$ the three lowest states remain
separated from the excited spectrum throughout the flux cycle, and the
insets show that they are permuted cyclically among themselves,
returning to the initial configuration only after three flux quanta, as expected for a $\nu = 1/3$ topological multiplet.

\begin{figure}
    \centering
    \includegraphics[width=0.3\linewidth]{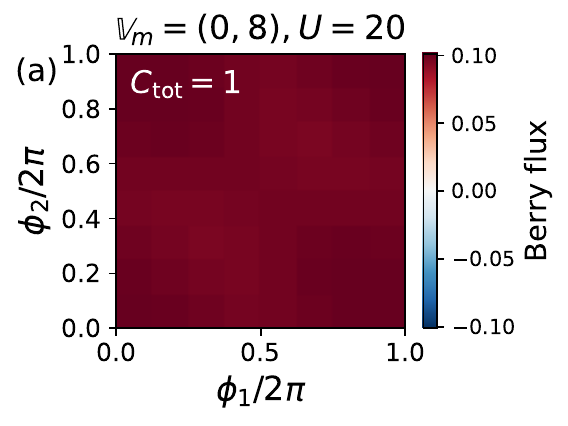}
    \includegraphics[width=0.3\linewidth]{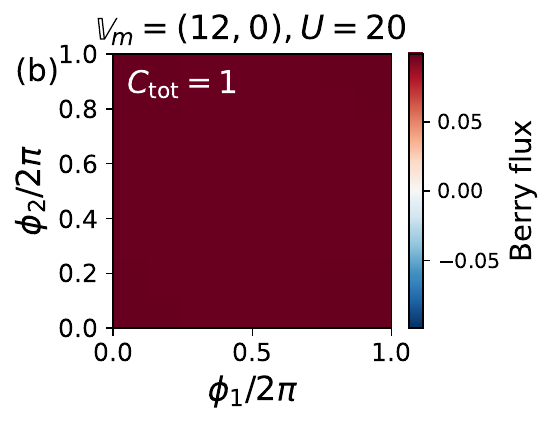}
    \caption{Many-body Berry curvature of the ground state manifold for $N_s=36, N_e=4$, and (a) $\mathbb V_m=(0,8), U=20$; (b) $\mathbb V_m=(12,0), U=20$. The color scale shows the Berry flux in each plaquette on an $8\times 8$ $\phi$-grid. The total many-body Chern number is $C_{\rm tot}=1$ for both (a) and (b).}
    \label{fig:berry_FCI}
\end{figure}

Fig.~\ref{fig:berry_FCI} shows the many-body Berry curvature of the
threefold manifold on an $8\times8$ twist grid, computed from
Eqs.~\eqref{eq:SM_berryflux} and~\eqref{eq:SM_chern}. The flux is
positive and nearly uniform on every plaquette.
Summing gives $C_{\rm tot} = 1$ for both geometries, i.e.\
$\sigma_{xy} = -e^2/3h$.

\begin{figure}
    \centering
    \includegraphics[width=0.495\linewidth]{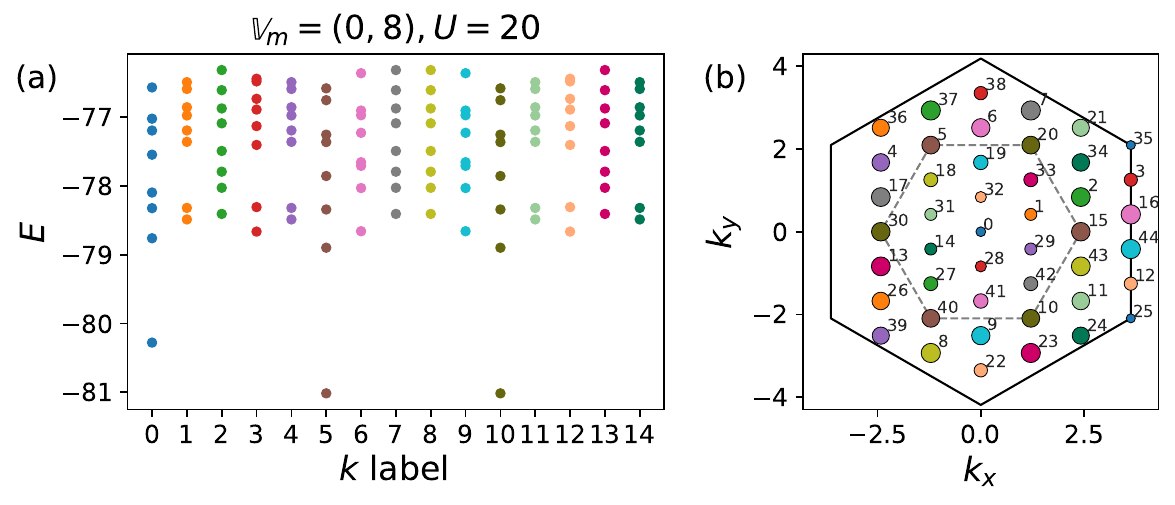}
    \includegraphics[width=0.495\linewidth]{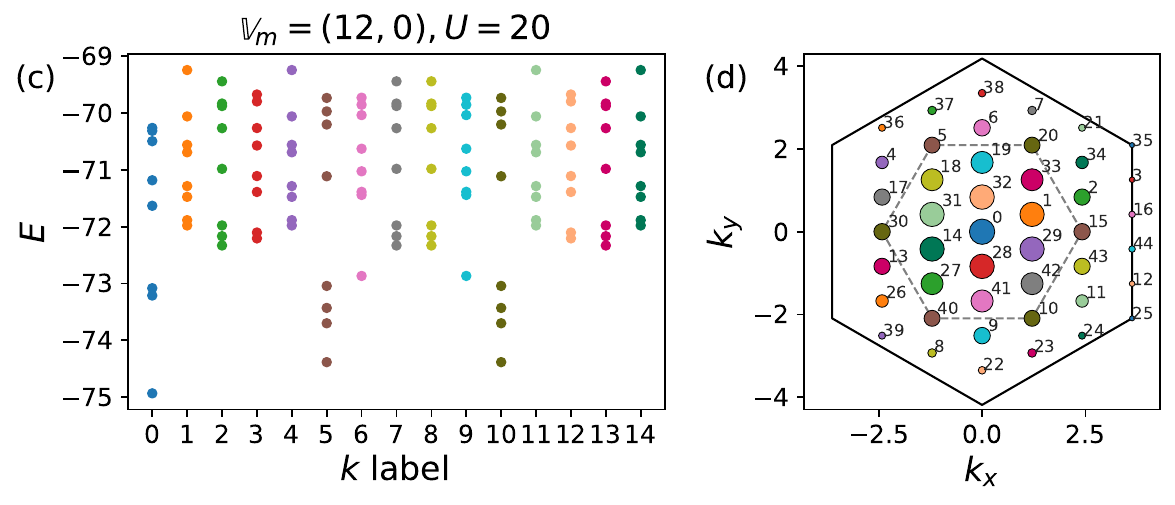}
    \caption{Many-body energy spectrum and Bloch state occupation (averaged over the lowest three states, indicated by the size of $\bm k$-points in b, d) for $N_s=45, N_e=5$, and (a, b) $\mathbb V_m=(0,8), U=20$; (c, d) $\mathbb V_m=(12,0), U=20$.}
    \label{fig:FCI_Ns45}
\end{figure}

\begin{figure}
    \centering
    \includegraphics[width=0.495\linewidth]{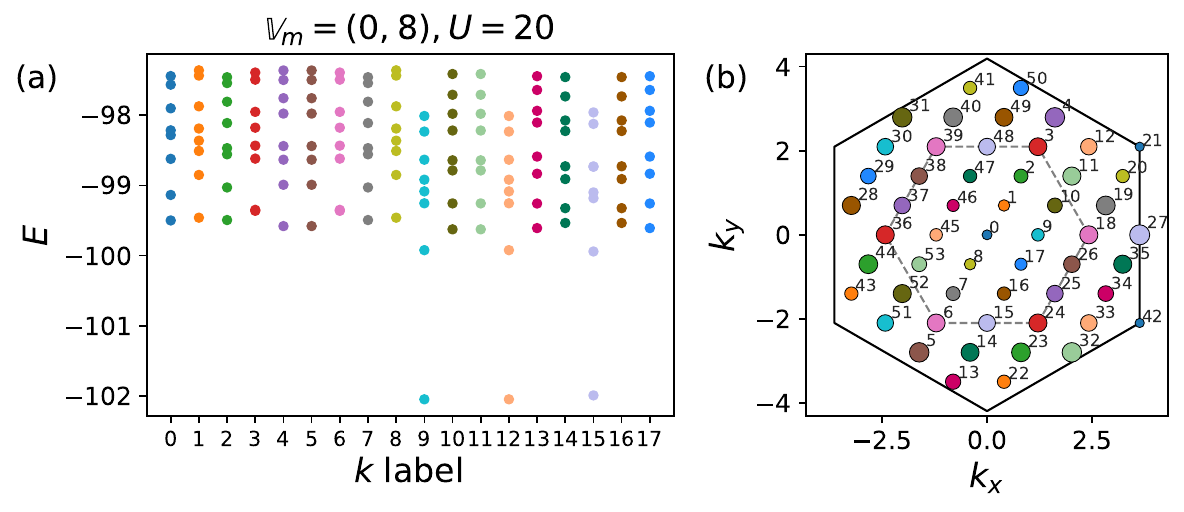}
    \includegraphics[width=0.495\linewidth]{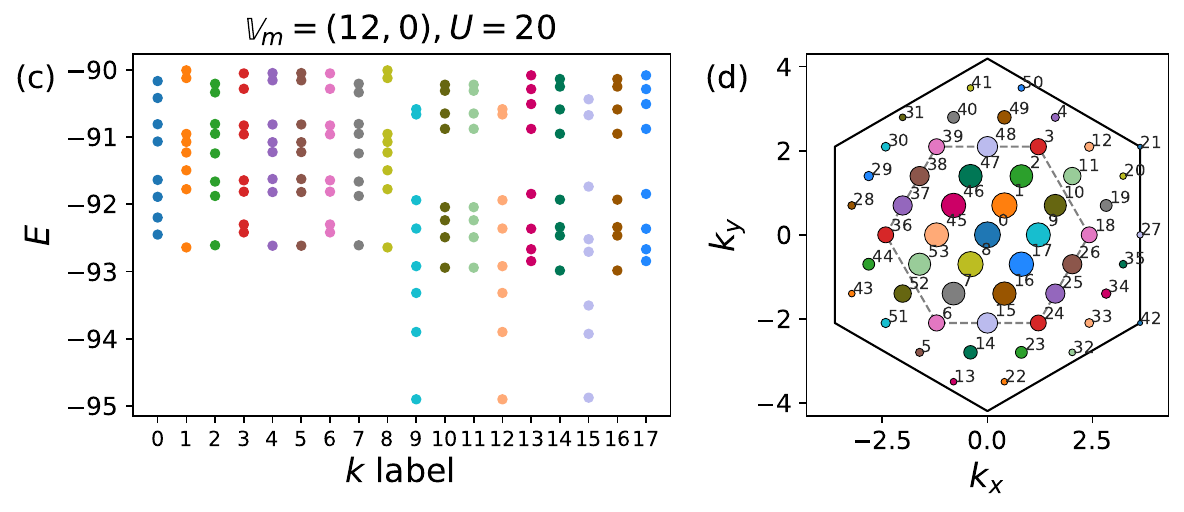}
    \caption{Many-body energy spectrum and Bloch state occupation (averaged over the lowest three states, indicated by the size of $\bm k$-points in b, d) for $N_s=54, N_e=6$, and (a, b) $\mathbb V_m=(0,8), U=20$; (c, d) $\mathbb V_m=(12,0), U=20$.}
    \label{fig:FCI_Ns54}
\end{figure}

Figs.~\ref{fig:FCI_Ns45} and~\ref{fig:FCI_Ns54} show the many-body spectra and
Bloch-state occupations for $N_s = 45$, $N_e = 5$ and $N_s = 54$,
$N_e = 6$, at both $\mathbb V_m = (0,8)$ and $(12,0)$. In all cases a threefold multiplet is
separated from the excited spectrum by a clear gap.
The splitting within the multiplet is not monotonic in system size: it
is small at $N_s = 36$ and $54$ but appreciably larger at $N_s = 45$,
where the three states are visibly non-degenerate on the scale of the
gap. We attribute this to a cluster-dependent finite-size effect.
Importantly, the diagnostics (see below) used to identify the phase do not require
the multiplet to be degenerate, only that it be isolated from the
excited spectrum.
The Bloch-state occupations, averaged over the three ground states,
reproduce the pattern of the main text at every size: the occupation is
enhanced at the momenta where the basin band is polarized to the layer
on which $V_m$ acts---at large $|\bm k|$ for
$\mathbb V_m = (0,8)$ and near the band bottom for $(12,0)$.

\begin{figure}
    \centering
    \includegraphics[width=0.9\linewidth]{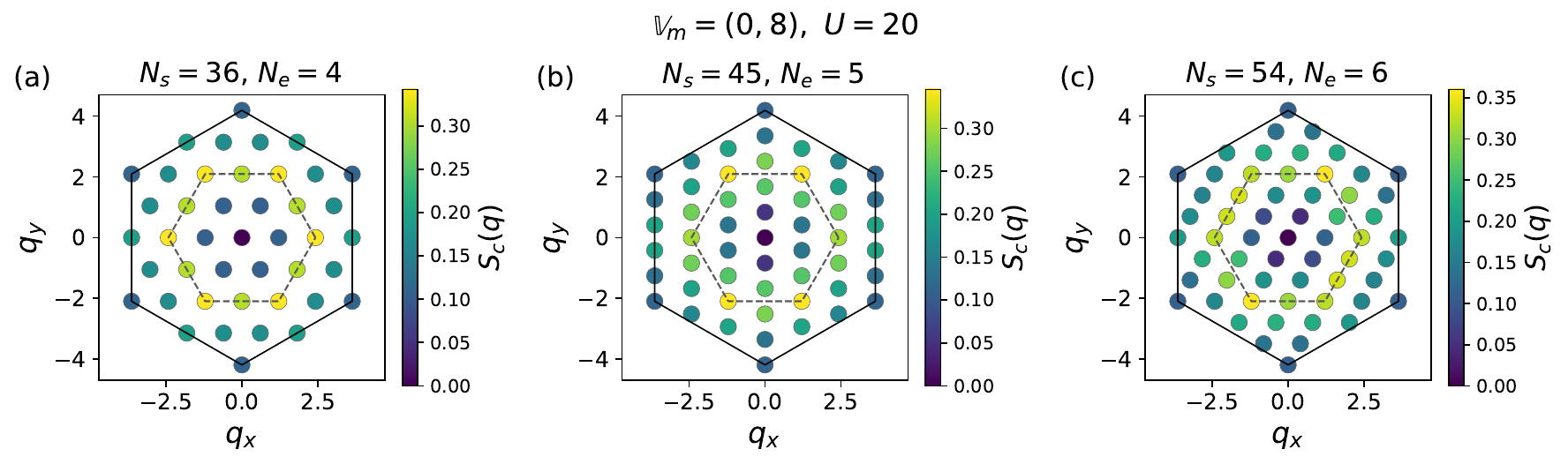}
    \includegraphics[width=0.9\linewidth]{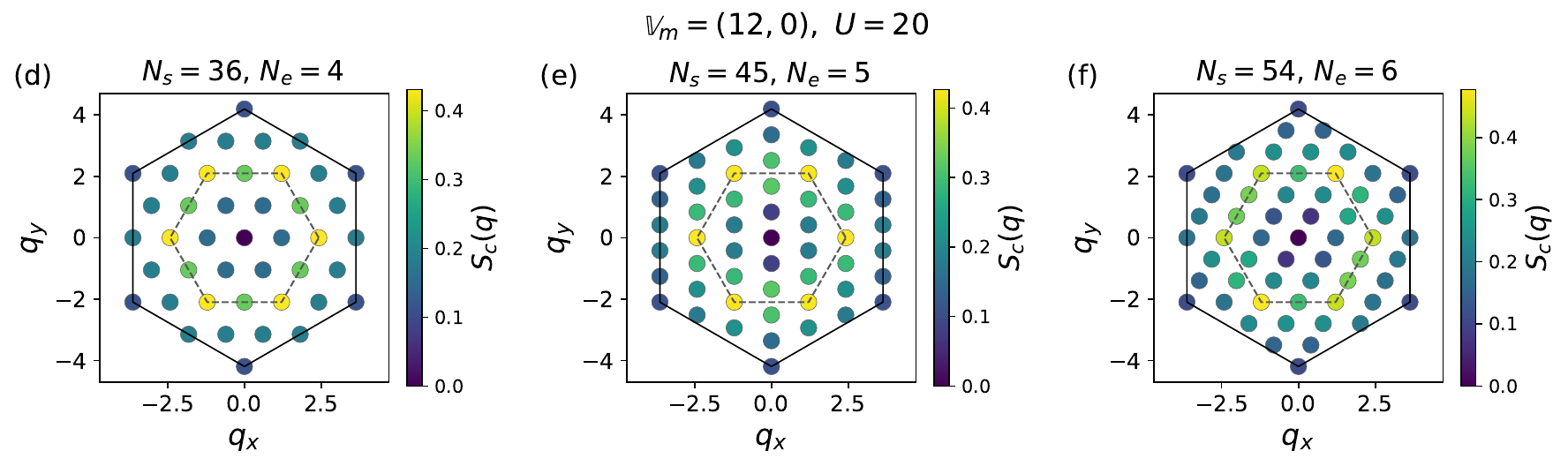}
    \caption{Structure factor $S_c(\bm q)$ averaged over the three ground states at $U=20$, (a, b, c) $\mathbb V_m=(0,8)$; (d, e, f) $\mathbb V_m=(12,0)$, for three system sizes: (a, d) $N_s=36, N_e=4$; (b, e) $N_s=45, N_e=5$; (c, f) $N_s=54, N_e=6$.}
    \label{fig:Sq_FCI}
\end{figure}

Fig.~\ref{fig:Sq_FCI} shows the connected structure factor averaged over
the three ground states for $N_s = 36$, $45$, and $54$ for both
$\mathbb V_m$. The maxima lie at the corners $\pm\bm\kappa$ of the mBZ,
which are wavevectors that correspond to a further tripling of the moir\'e cell.
These peaks nonetheless reflect correlations rather than order. The
peak height in $S_c$, already normalized by $N_e$, stays nearly constant with increasing system size: $S_c(\pm\bm\kappa) \simeq 0.342, 0.330, 0.336$ for $N_e=4,5,6$ (for non-$C_3$-symmetric clusters, this is the average $S_c$ over $C_3$-related $\bm\kappa$ points). This is
in stark contrast to the moir\'eless crystal in Fig.~\ref{fig:Sq_AHC}, where
the corresponding peak grows linearly with $N_e$.

\begin{figure}
    \centering
    \includegraphics[width=0.9\linewidth]{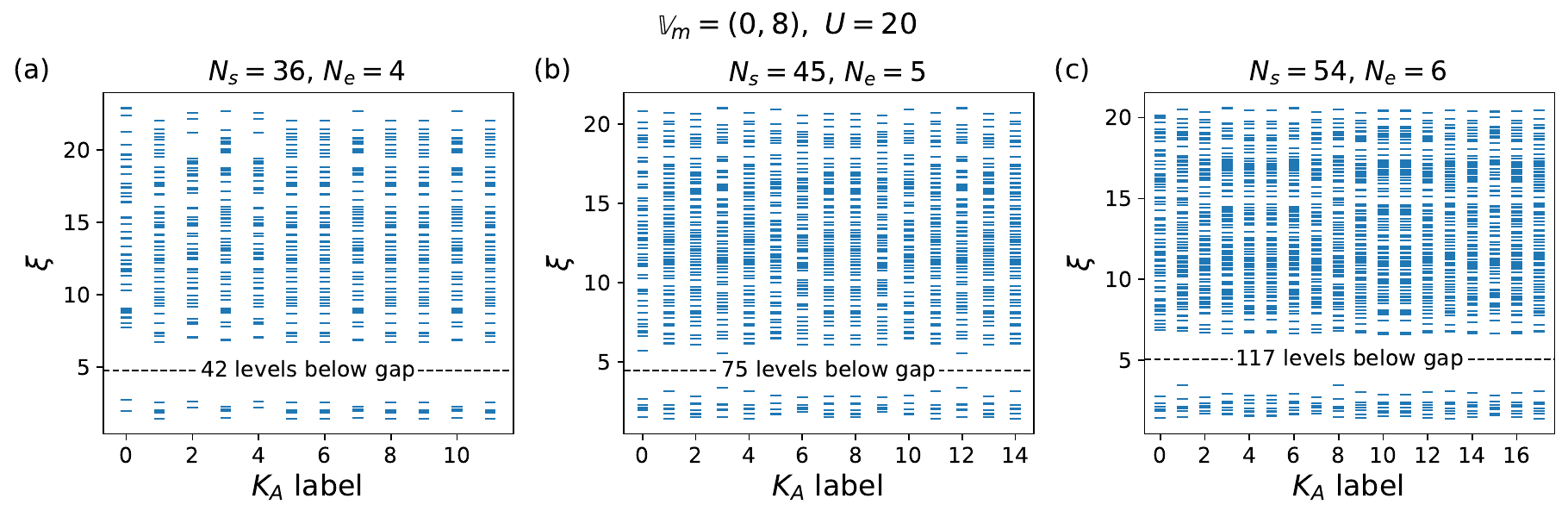}
    \includegraphics[width=0.9\linewidth]{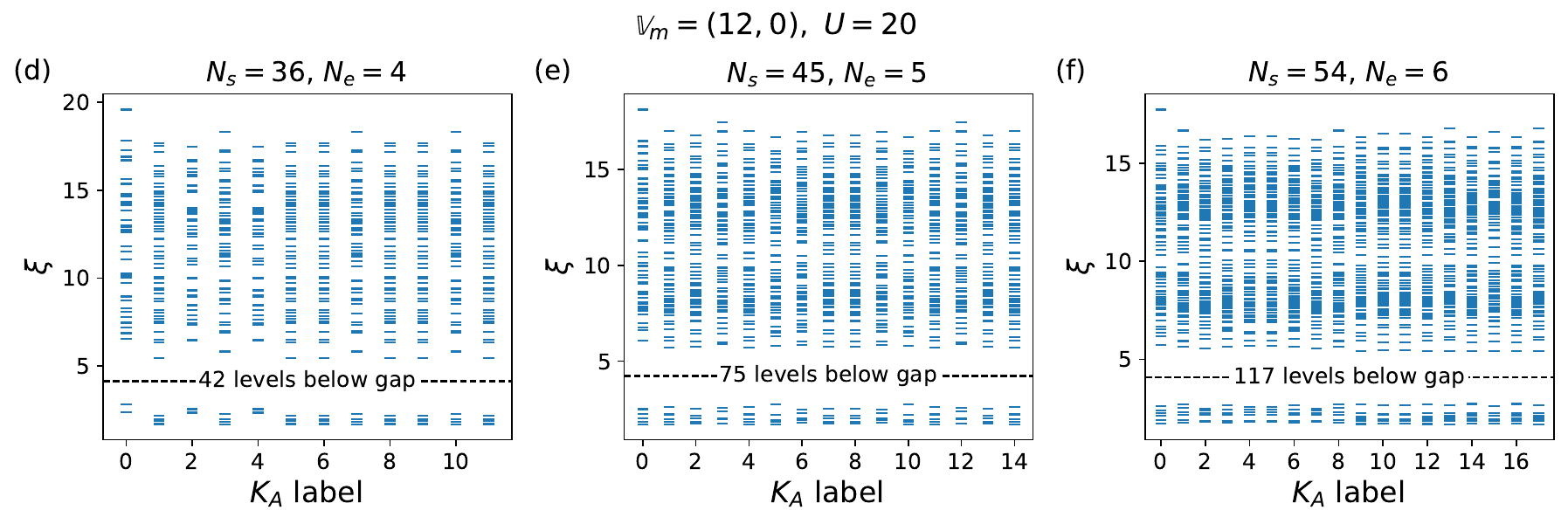}
    \caption{Particle entanglement spectrum for $U=20$, (a, b, c) $\mathbb V_m=(0,8)$; (d, e, f) $\mathbb V_m=(12,0)$, for three system sizes: (a, d) $N_s=36, N_e=4$; (b, e) $N_s=45, N_e=5$; (c, f) $N_s=54, N_e=6$.}
    \label{fig:PES}
\end{figure}

An important test of Laughlin-type topological order is the counting of
levels in the particle entanglement spectrum
(PES)~\cite{regnault2011fractional}. Tracing
the density matrix of the threefold ground-state manifold over
$N_e - N_A$ particles and diagonalizing, one obtains an entanglement
spectrum $\xi$ resolved by the momentum $K_A$ of the retained
particles. For a Laughlin state the number of levels below the
entanglement gap equals the number of quasihole states of $N_A$
particles on $N_m$ orbitals (in our model $N_m=N_s/3$), given by the $(1,3)$-admissible counting
\begin{equation}
\mathcal N(N_A, N_m) = \frac{N_m}{N_m - 2N_A}
\binom{N_m - 2N_A}{N_A} .
\label{eq:SM_pes_counting}
\end{equation}
Fig.~\ref{fig:PES} shows the PES for $N_A = 2$ at the three system
sizes and both sets of parameters. The counting is reproduced exactly:
\begin{center}
\begin{tabular}{cccc}
\hline\hline
$N_s$ & $N_m$ & $\mathcal N(2,N_m)$ & PES \\
\hline
$36$ & $12$ & $42$  & $42$  \\
$45$ & $15$ & $75$  & $75$  \\
$54$ & $18$ & $117$ & $117$ \\
\hline\hline
\end{tabular}
\end{center}
in all six panels.
The agreement holds for both the moir\'e-distant and moir\'e-proximate
geometries and at all three sizes, establishing that the ground state
carries the topological order of the $\nu = 1/3$ Laughlin state.

\section{Triangular moir\'e potential}

For $V_m < 0$ the potential minima form a triangular rather than a
honeycomb lattice, with a single minimum per superlattice cell. This
changes the single-particle problem qualitatively, and the resulting
phase diagrams differ correspondingly from those of the main text.

\begin{figure}
    \centering
    \includegraphics[width=0.7\linewidth]{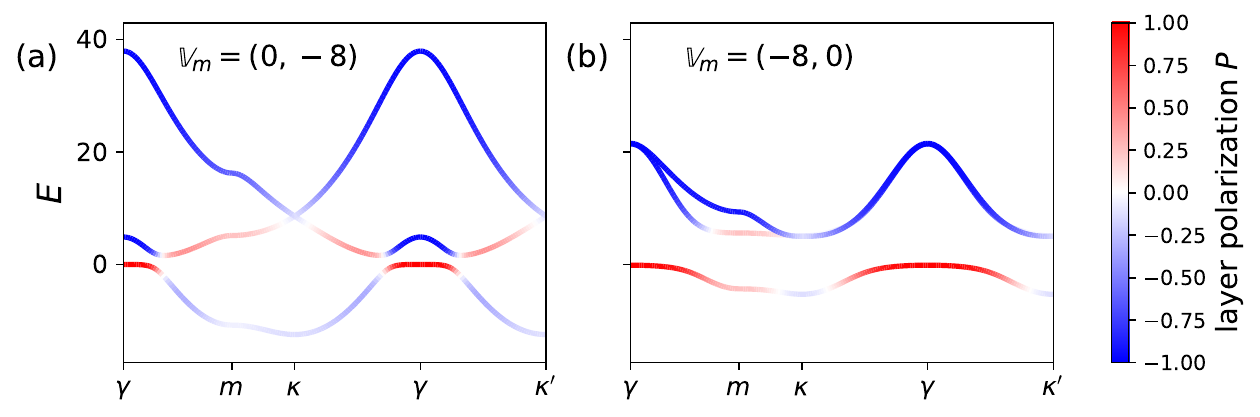}
    \caption{Band structure of the Berry basin model with $N=3,\lambda=1,\beta=0.2,V_0=-25$, and triangular moir\'e potential (a) $\mathbb V_m=(0,-8)$; (b) $\mathbb V_m=(-8,0)$.}
    \label{fig:bands_triang}
\end{figure}

Fig.~\ref{fig:bands_triang} shows the minibands for $V_m = -8$ on the top or bottom
layer. In contrast to the honeycomb case, the triangular potential
separates the lowest two minibands: with a single minimum per
superlattice cell the crossings at $\kappa$ and $\kappa'$ are no longer
protected, and the lowest miniband is gapped from higher bands across the whole mBZ.
A well-defined noninteracting ground state at $\nu = 1$ therefore
exists here, unlike for $V_m > 0$.

When the potential acts on the bottom layer [Fig.~\ref{fig:bands_triang}(a)], the gap between the lowest two minibands is small because the Bloch states at the basin bottom are strongly polarized to the top layer. At small $|V_m|$ the lowest miniband carries Chern number $C = 3$; at $|V_m| \simeq 10$ the single-particle gap closes at $\gamma$ and
reopens with the lowest miniband trivial, $C = 0$.
For the potential on the top layer [Fig.~\ref{fig:bands_triang}(b)], the
lowest miniband is gapped and topologically trivial at all $|V_m|>0$.

\begin{figure}
    \centering
    \includegraphics[width=\linewidth]{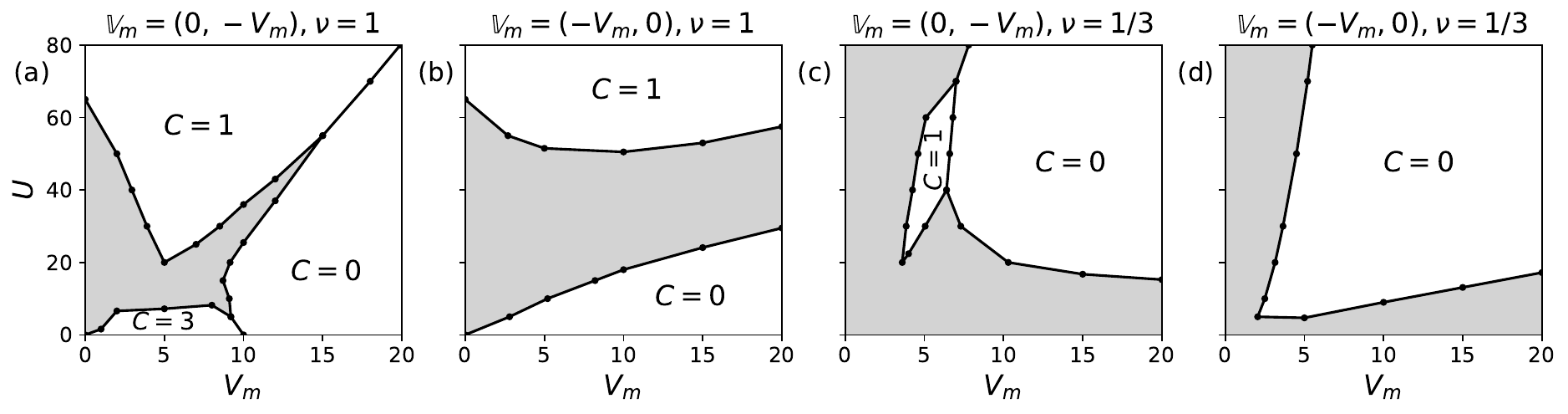}
    \caption{Phase diagrams for triangular moir\'e potential $-V_m<0$ at fillings (a, b) $\nu=1$ and (c, d) $\nu=1/3$. The gray and white regions represent metallic states and insulating states with Chern number $C$.}
    \label{fig:phase_triang}
\end{figure}

Figs.~\ref{fig:phase_triang}(a) and (b) show the phase diagrams at moir\'e
filling $\nu = 1$. The weak-coupling structure follows directly from
the miniband topology: the $C = 3$ region at $U \lesssim 8$,
$V_m \lesssim 10$ in (a) is the noninteracting state obtained by
filling the lowest miniband, and its boundary to the $C = 0$ region is
the single-particle gap closing described above. In (b) the weak-coupling state is trivial throughout,
again as expected from the band topology.

At large $U$ a $C = 1$ Chern insulator appears in both (a) and (b).
Its Chern number matches neither miniband value, and it connects
continuously to the moir\'e\-less anomalous Hall crystal at $V_m = 0$,
so it is interaction-driven in the same sense as the states discussed
in the main text. In the ED spectrum this state is two-fold degenerate because of its sublattice structure: the anomalous Hall crystal
has two charge density maxima per unit cell, forming a honeycomb
lattice~\cite{zeng2024sublattice}, whereas the triangular moir\'e
potential has a single minimum per cell. Only one of the two maxima can
be pinned to that minimum, and the two choices are related by the
sublattice exchange that acts as inversion on the honeycomb crystal.
The moir\'e potential therefore selects between two
inversion-related configurations rather than a unique one, leaving a
two-fold ground-state manifold with $C_{\rm tot} = 2$.

At $\nu = 1/3$ no fractional state appears when the triangular potential acts on either layer
[Figs.~\ref{fig:phase_triang}(c) and (d)]. For the bottom layer a narrow
$C = 1$ region survives near $V_m \sim 5$, bounded both above
and below; elsewhere, and everywhere in (d), the ground state is either metallic or
a trivial $C = 0$ insulator. Both the $C=0$ and $C=1$ insulators at $\nu=1/3$ break the moir\'e translation symmetry down to that of a $\sqrt{3}\times\sqrt{3}$ supercell; the former is a trivial charge density wave while the latter is a commensurate anomalous Hall crystal.

\end{document}